\documentclass[11pt]{article}

\usepackage{amsmath}
\usepackage{amssymb}

\usepackage{mathpazo}

\usepackage{graphicx}

\usepackage{cite}

\usepackage{color} 

\usepackage{setspace} 
\usepackage{dcolumn}
\newcolumntype{.}{D{.}{.}{-1}}

\usepackage{marginnote}
\newcommand{\mycomment}[1]{}

\usepackage[a4paper,margin=3cm]{geometry}
\usepackage{url}
\usepackage[labelfont=bf,labelsep=period,justification=raggedright]{caption}

\usepackage{lastpage}

\usepackage{namedplus}
\usepackage{xspace}

\emergencystretch=\maxdimen
\date{}

\providecommand*{\ap}[1]{%
\ensuremath{^\mathrm{#1}}}
\newcommand{\gMathV}{\textit{Math5}$\ap{\mathrm{-/-}}$\xspace}

\newcommand{\gIsl}{\textit{Isl2}\xspace}
\newcommand{\gIslE}{\textit{Isl2-EphA3}\xspace}
\newcommand{\gIslEkk}{\textit{Isl2-EphA3}$\ap{\mathrm{ki/ki}}$\xspace}
\newcommand{\gIslEkp}{\textit{Isl2-EphA3}$\ap{\mathrm{ki/+}}$\xspace}

\newcommand{\gIslp}{\textrm{Isl2\ap{+}}\xspace}
\newcommand{\gIslm}{\textrm{Isl2\ap{-}}\xspace}

\newcommand{\gTKO}{\textit{{ephrin-A2,A3,A5}}\xspace}

\newcommand{\gEphAIII}{\textit{EphA3}\xspace}
\newcommand{\gEphAIV}{\textit{EphA4}\xspace}
\newcommand{\gEphAV}{\textit{EphA5}\xspace}

\usepackage{upgreek}
\newcommand{\betaIIKO}{\ensuremath{\beta\mathit{2}^{-/-}}\xspace}

\begin{document}

\singlespacing

\noindent \today
\begin{flushleft}
{\Large
\textbf{Quantitative assessment of computational models for
  retinotopic map formation}
}\\
\vspace{0.2cm}
Abbreviated title: Quantitative assessment of retinotopic map models\\
\vspace{0.5cm}
J J Johannes Hjorth$^{1,\ast}$, 
David C Sterratt$^{2}$, 
Catherine S Cutts$^{1}$,
David J Willshaw$^{2}$, 
Stephen J Eglen$^{1}$
\\
\vspace{0.5cm}
1 Cambridge Computational Biology Institute, Department of Applied Mathematics and Theoretical Physics, University of Cambridge, Cambridge CB3 0WA, UK
\\
\vspace{0.1cm}
2 Institute for Adaptive and Neural Computation, School of Informatics, University of Edinburgh, 10 Crichton Street, Edinburgh EH8 9AB, UK
\\
\vspace{0.1cm}
$^\ast$ Corresponding author: j.j.j.hjorth@damtp.cam.ac.uk
\\
\end{flushleft}
Number of pages: \textbf{\pageref{LastPage}} \\
Number of figures: \textbf{11} \\
Number of tables: \textbf{5} \\
Number of words for: Abstract (\textbf{231}, max 250) \vspace{0.2cm}\\
\vspace{0.5cm}

Keywords: mouse,
retinocollicular projection,
retinotopic map formation,
computational modelling framework,
quantitative evaluation\\

\vspace{0.5cm}

Conflict of interest: The authors
  declare no competing financial interests.\\
  \subsubsection*{Acknowledgements}
  Experimental data were provided by Daniel Lyngholm and Ian Thompson
  (wild type development), Michael Reber and Greg Lemke (anterograde
  injection data and retinal EphA gradients), Jianhua Cang (intrinsic
  imaging data). The authors thank Alexei Koulakov for providing his
  model's code for reference and David Nissenbaum for sharing his
  Gierer 2D code with the project.  We thank Kate Belger for
  proofreading.  The authors thank the Wellcome Trust (JJJH, DCS, DJW,
  SJE; grant number 083205) and EPSRC (CSC) for
  funding.\\

  \subsubsection*{Contributions}
  Designed research, performed research, analysed data, wrote the
  paper: JJJH, DCS, CSC, DJW, SJE.

\subsubsection*{Abbreviations}

\begin{tabular}{ll}
  \hline
RGC  & Retinal Ganglion Cell\\
SC & Superior Colliculus\\
TKO  & \textit{Ephrin-A2,A3,A5} Triple Knock-Out\\
\hline
\end{tabular}

\clearpage

\onehalfspacing

\section*{\uppercase{Abstract}}

Molecular and activity-based cues acting together are thought to guide
retinal axons to their terminal sites in vertebrate optic tectum or
superior colliculus to form an ordered map of connections. The details
of mechanisms involved, and the degree to which they might interact,
are still not well understood. We have developed a framework within
which existing computational models can be assessed in an unbiased and
quantitative manner against a set of experimental data curated from
the mouse retinocollicular system. Our framework facilitates
comparison between models, testing new models against known phenotypes
and simulating new phenotypes in existing models. We have used this
framework to assess four representative models that combine Eph/ephrin
gradients and/or activity-based mechanisms and competition. Two of the
models were updated from their original form to fit into our
framework.  The models were tested against five different phenotypes:
wild type, \gIslEkk, \gIslEkp, \gTKO triple knock-out and \gMathV
(\textit{Atoh7}). Only one model could account for the collapse point
in \gIslEkp, and two models successfully reproduced the extent of the
\gMathV anteromedial projection. The models needed a weak
anteroposterior gradient in the superior colliculus in order to
reproduce the residual order in the \gTKO triple knock-out phenotype,
suggesting either an incomplete knock-out or the presence of another
guidance molecule.  This points to the need for future
modelling and experimental work to improve our understanding of the
developmental mechanisms involved.

\clearpage
\section*{\uppercase{Introduction}}

\begin{figure}[!ht]
\begin{center}
\includegraphics[width=10cm]{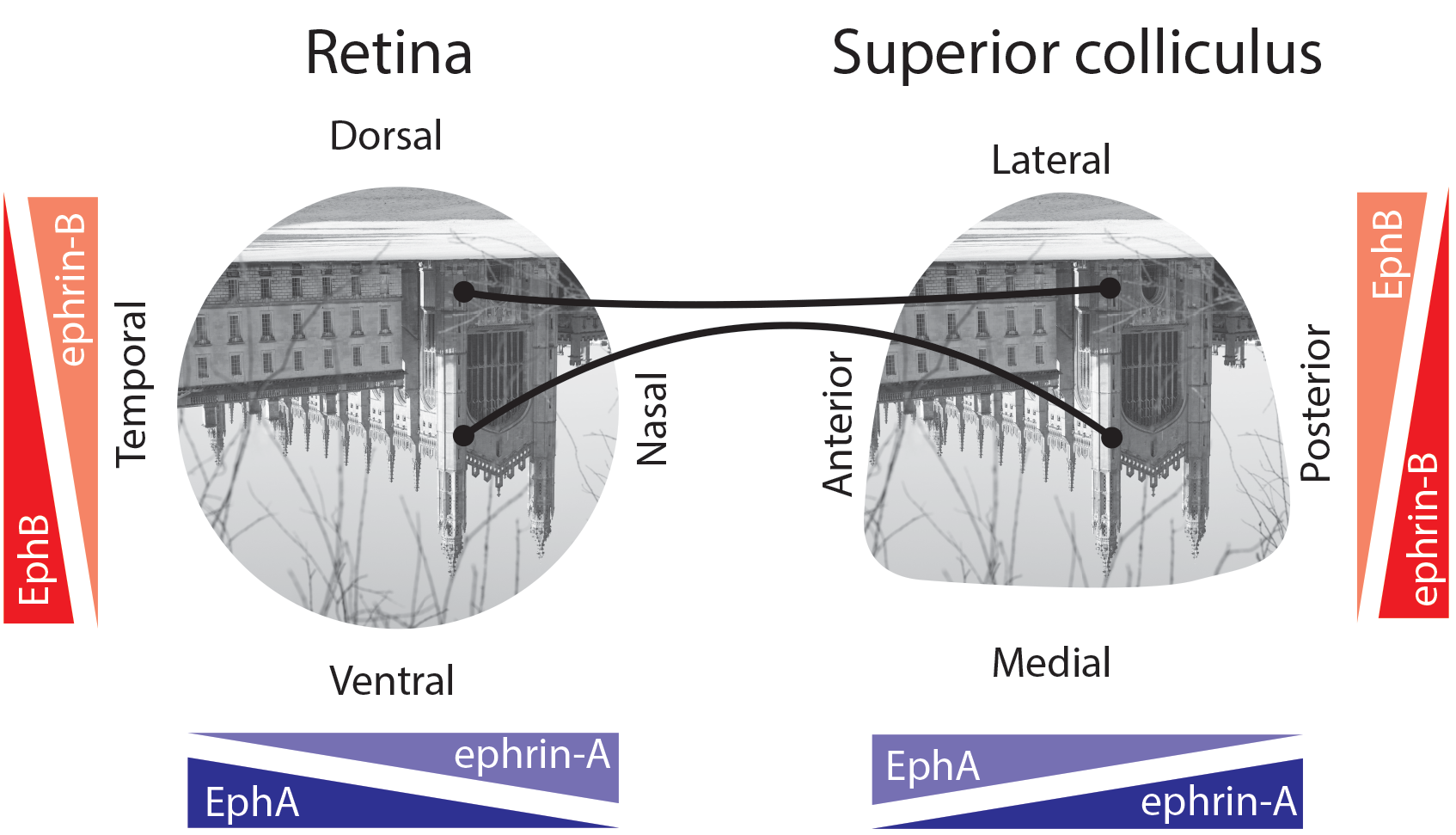} 
\end{center}
\caption{ {\bf Schematic of retinotopic map formation.} Retinal
  neurons project to SC in a topographic fashion. Each axis has an
  independent set of gradients instructing the map formation. Eph
  receptors of two different families are expressed across the retina
  in a graded fashion. The retinal EphA receptor gradient is low
  nasally and high temporally, whereas the retinal ephrin-A ligand
  countergradient has the opposite direction. In the SC, the ephrin-A
  ligand gradient goes from low anterior to high posterior, while the
  EphA receptor countergradient in the SC is in the opposite
  direction. The retinal EphB receptor gradient goes from ventral
  (high) to dorsal (low), while the ephrin-B ligand countergradient is
  in the opposite direction \protect \cite{McLaughlin2005}. In the
  SC, the ephrin-B ligand gradient goes medial (high) to lateral
  (low), and the EphB receptor countergradient has the reverse slope
  \protect \cite{Hindges2002,McLaughlin2005}.}
\label{fig:mapexplained}
\end{figure}

Many sensory systems are organised into topographic maps, where
neighbouring neurons in the source structure project to neighbouring
neurons in the target structure \cite{Cang2013review}. The mechanisms
involved in generating sensory maps may also be involved in the
development of other systems \cite{Cang2013review}. The mouse
retinotopic map (Figure~\ref{fig:mapexplained}) provides a model
system to study topographic map formation, with an extensive range of
mutant mice lines available
\cite{Frisen1998,Brown1998,Brown2000,Feldheim2000,Triplett2011}. During
development, axons from retinal ganglion cells (RGCs) grow through the
optic tract to innervate the superior colliculus (SC). By postnatal
day 1 (P1) RGC axons have grown all the way from the anterior to the
posterior region of the SC, overshooting their final target locations
\cite{McLaughlin2003}. The axons start branching, and then branches
outside the topographically correct location are pruned away
\cite{McLaughlin2003c}. The map is topographically ordered before eye
opening at P10--12 \cite{McLaughlin2003}, but the axonal arbor size
continues to decrease for a few more weeks \cite{Lyngholm2013}.

Several candidate mechanisms have been proposed 
to guide RGC axons to their final locations: 1. The chemoaffinity
hypothesis \cite{Sperry1963,Meyer1998} which in its modern form have Ephs
and ephrins labelling orthogonal axes in the retina and SC
\cite{McLaughlin2005}.  2. Spontaneous activity in the retina
which instructs map formation \cite{Ackman2012} via Hebbian-based
modification of synaptic strengths \cite{Willshaw1976}. 3. Competition
for resources or space in the target tissue
\cite{Triplett2011,Ooyen2011}.  4. Partial mediolateral ordering of
RGC axons within the optic tract \cite{Plas2005}.  5. Axon-axon
interactions \cite{Yates2004,Gebhardt2012}.

Insights from experiments with mutant mice gave rise to new computer
models, several of which have been reviewed
\cite{Swindale1996review,Goodhill1999review,Goodhill2005review,Goodhill2007review}.
However these reviews were qualitative and exclude recent genotypes
\cite{Cang2008,Triplett2011}. We have created an open framework to
compare model results quantitatively with experimental data and
compare models with each other.

We aimed to see if any model, under one set of parameter values, is
consistent with all phenotypes.  To make the task tractable we
reimplemented a representative subset of models
\cite{Whitelaw1981,Gierer1983,Willshaw2006,Triplett2011} and applied
them to phenotypes previously described in sufficient quantitative
detail \cite{Feldheim2000,Reber2004,Cang2008,Triplett2011}. Key
features of the resulting maps are quantified using virtual
experiments and compared to experimental data. Our findings suggest
that the models failed to account for the range of experimental data
studied. Only one model can reproduce the collapse point seen in the
\gIslEkp phenotype, and two of the models fail to reproduce the \gMathV
phenotype. However, by reintroducing a weak gradient in the SC the
models can reproduce the global order still remaining in \gTKO triple knock-out maps.

\clearpage

\section*{\uppercase{Methods}}

The modelling process had three main stages: (i)~selection of mouse
genotypes with retinotopic map data; (ii)~selection of models from the
literature to test against the data; and (iii)~simulation of these
models and comparison with the data.  To enable a precise,
quantitative comparison between different models and to generate the
predictions, we simulated all models within the same modelling
pipeline.  The model pipeline has three phases comprising calculation
of initial conditions, simulation of the development of connections,
and analysis of the final connection patterns.  All computer code and
data relating to this project (pipeline, models, analysis tools) are
freely available
(\mbox{\url{https://github.com/Hjorthmedh/RetinalMap}}).

\subsection*{Genotype selection}

We used experiments from five mouse genotypes for which we believe
there are sufficient quantitative data to constrain the models and
which are important in ruling out certain classes of model.

1. The most quantitative information comes from wild type mice, with
both anatomical tracing data across development \cite{Lyngholm2013},
and whole maps acquired by intrinsic imaging data from adult mice
\cite{Cang2008}.

2. The \gIslE genotypes (heterozygous and homozygous knock-in)
disrupts the molecular positional information for around $40\,\%$ of
the RGCs by adding extra EphA3, providing phenotypes where we can
assess the impact of systematically modifying gradients upon maps. The
phenotypes from \gIslE are characterised along projections from
nasotemporal (NT) axis to the anteroposterior (AP) axis using retinal
injections \cite{Brown2000,Reber2004}. Further combinations of \gIslE
with \gEphAIV and \gEphAV knock-outs \cite{Reber2004,Bevins2011} were
analysed, but omitted here as results were qualitatively similar to
earlier findings \cite{Willshaw2006}.

3. In triple knock-out (TKO) of \gTKO, all the ephrin-As participating
in map formation along the anteroposterior axis of the SC were
removed. The whole map has been characterised by intrinsic imaging
\cite{Cang2008} and analysed using the Lattice method
\cite{Willshaw2014}.

4. The \gMathV knock-out has a reduced RGC population in the retina,
reducing competition between RGC axons. The phenotype has been
characterised mainly by whole eye injections that give the density of
the SC projections \cite{Triplett2011}.

Many other mutant mice lines have been characterised by antereograde
or retrograde labelling of axons, including knockouts of ephrin-A2 and
ephrin-A5 \cite{Feldheim2000} and EphA7 \cite{Rashid2005}. This data
is more challenging to quantify as (i)~there is one injection site per
individual and (ii)~there appears to be considerable variation in the
locations of termination zones between individuals
\cite{Feldheim2000}. The variability means it is not possible to
create a single composite map (as in the case of the \gIslE knock-ins)
from multiple individuals. We therefore decided to exclude these data
from this quantitative comparison.  We also excluded mutant mice lines
that perturbed retinal activity (e.g. \citetext{McLaughlin2003}) as
two of the models studied here exclude activity-dependent mechanisms.

\subsection*{Choice of models}

The main criteria used for our choice were that (i) the models
contained mechanisms providing for flexibility in the pattern of
connections formed; (ii)~the models simulated the development of two
dimensional maps or could be extended so to do and (iii)~they had
explicit representations of gradients to allow manipulations in
gradients to be simulated.

(i) \citetext{Prestige1975} suggested a classification of the different
ways in which graded labels could instruct retinotopic mappings. In
Type I mechanisms, gradients provide the highest affinity for the
correct location \cite{Sperry1963,Meyer1998}. In Type II mechanisms,
all axons prefer the same location, but with different affinity
\cite{Prestige1975}.  Together with a competition mechanism the map
then organises itself so that the RGC with highest affinity for the
location with highest affinity innervates it, leaving the next most
affine RGC to innervate the next most affine SC neuron, and so on.
Type~I models establish connections by matching up fixed-value labels
on RGC axons with those on SC neurons. In the \gIslE mutant maps the
abnormally high values of EphA in much of the retina have no
counterpart in the colliculus yet all the retina projects to the
colliculus. This finding rules out strict Type~I models.

(ii) We excluded the 1D branching model due to \citetext{Yates2004} as
we were unable to make a 2D model from the information provided. 

(iii) We also excluded the model of \citetext{Simpson2011} as
chemoaffinity is represented implicitly, by a term describing the
distance of an axon from its correct location, and the model by
\citetext{Grimbert2012} as no method was given to convert gradients to
their probability maps used in their simulations \cite{Sterratt2013c}

We selected four models that include a range of developmental
mechanisms implicated in the development of retinotopic maps.
\cite{Sterratt2013c}.  Here we refer to the models by the surname of
either the first author of the relevant publication or the principal
architect. We chose the following models:
\begin{enumerate}
\item The \citetext{Gierer1983} model exists as both Type~I and
  Type~II versions, the Type~II version including a mechanism akin to
  competition. Here we use an updated version of Gierer's Type~II
  model \cite{Sterratt2013b} in which the strength of competition can
  be modified.

\item The Koulakov model \cite{Triplett2011} is a generalisation of
  the Gierer model including an abstract representation of correlated
  retinal activity.

\item The Whitelaw model \cite{Whitelaw1981} combines a Hebbian
  activity scheme \cite{Willshaw1976} with a Type~II affinity
  mechanism. It has an explicit representation of retinal activity and
  a multiplicative interaction between activity and gradients.

\item The Willshaw model \cite{Malsburg1977,Willshaw2006}, also known
  as the ``Marker Induction model'', uses a Type~I gradient matching
  scheme where the SC gradients are modifiable during development by
  the action of the incoming retinal fibres.
\end{enumerate}

The Gierer, Whitelaw and Willshaw models were proposed before the
discovery of Ephs and ephrins \cite{Gierer1983,Whitelaw1981,Malsburg1977}.
In later versions of both the Gierer model and the Willshaw model the
specifics of these graded labels were introduced
\cite{Sterratt2013b,Willshaw2006}. Here we have made additional
extensions to make all models two-dimensional.  In all cases a single
molecule type (A or B) labels each axis of the retina and the SC.  The
Gierer model has spatially-restricted sprouting, such that new
synapses are generated close to existing ones (as did the original
Willshaw model \cite{Malsburg1977}); in the other models, new synapses
can be placed with fewer constraints in the SC, irrespective of the
location of previous synapses.

\subsection*{Pipeline phase 1 --- Initialisation}

The positions of neurons in retina and SC and the concentration of
EphA/B receptors and ephrin-A/B ligands define the initial conditions
of the simulations for the different genotypes. These can then be
passed to one of the models defined below, and the retinotopic map
formation simulated. The initial connections set up by each model are
described in the relevant sections below.

\paragraph{Number and placement of neurons}

In mouse, there are around 50,000 RGCs
\cite{Jeon1998,SalinasNavarro2009}, and an unknown number of SC
neurons. Here networks containing 2,000 retinal neurons
($N_\mathrm{R}$) and 2,000 SC neurons ($N_\mathrm{SC}$), were
simulated. We believe these populations to be large enough to
represent the system, without being too large to make the models too
demanding in computation time.  The positions of neurons are drawn
randomly from a uniform 2D distribution
\cite{Galli-Resta1997,Eglen2012}. If there are no other neurons within
a certain specified distance, this position is accepted. The algorithm
terminates when the required number $N$ neurons have been placed
within the structure, or $1,000 \cdot N$ positions have been rejected
in total. To minimise edge effects neurons are also placed outside the
target structure, but are not counted in the final population. This
prevents an artificial inflation of the density of neurons at the
edges. The retinal size was normalised to unit size and the retinal
neurons were placed within a circle of diameter 1. The shape of the SC
was taken from Figure~2 in \citetext{Drager1976}. The minimum distance
was set separately for retina ($d_\mathrm{R}$) and SC
($d_\mathrm{SC}$) so that 2,000 neurons would fit inside the space
available (Table~\ref{tab:Parameters}).

\paragraph{Specifying Gradients}

Despite their importance for map formation \cite{McLaughlin2005}, the
only quantitative measures of Eph/ephrin gradients is for retinal EphA
where mRNA levels were measured at P1 using \textit{in situ}
hybridisation along the nasotemporal axis \cite{Reber2004} and
modelled as shallow exponential gradients. By contrast, there is no
quantitative data for ephrin-As or the B-system \cite{Hindges2002} and
so we have assumed exponential profiles.  In parallel with these
forward gradients there is a second set of opposing countergradients
of Eph receptors in the SC and ephrin ligands in the retina
(Figure~\ref{fig:mapexplained}). These countergradients have not been
quantified in either the retina or the SC. Since recent work showed
that countergradients can be replaced by a competitive mechanism
\cite{Sterratt2013b}, we have focused on the forward system and
excluded countergradients.  Table~\ref{tab:gradients} describes how we
have quantified the gradients, which are displayed pictorially in
Figure~\ref{fig:gradients}.  The gradients are identical for all
repeats of a given genotype, but they are sampled at the neuron
locations, which vary between runs.  We assume the affinities of the
receptor subtypes are similar and the individual gradients are summed
to give the total expression of EphA, EphB, ephrin-A and ephrin-B at
any point in retina and SC \cite{Brown2000,Bevins2011}.

To explore the effect of a weak signalling molecule, for \gTKO triple
knock-out we introduced a weak gradient running along the rostrocaudal
axis of the SC with the same shape as the ephrin-A gradient assumed for
the wild type but with strength multiplied by a constant $K<1$ to
scale it down.

\paragraph*{Model configuration}

To ensure a fair comparison, all models were created with the same
spatial geometry in retina and SC.  The number of neurons in retina
and SC was also fixed, and neurons were positioned according to the
minimal spacing rules described earlier (for parameters see
Table~\ref{tab:Parameters}, top four rows).  All models were
restricted to use one set of parameter values for all genotypes
(Table~\ref{tab:Parameters}, remaining rows). The parameter
values in three models were optimised manually to fit one experimental
condition (Gierer was optimised for \gMathV, Koulakov for \gIslEkp and
Whitelaw for \gIslEkk).  The Willshaw model did not require any
additional parameter tuning beyond that presented in 2006.

\subsection*{Pipeline phase II --- Running the simulations}

Models written in MATLAB, R and C have been integrated into the
pipeline. Implementation details of each model are described
later. Each genotype is run ten times with different initial
conditions (positions of neurons, gradients and initial
connectivity) for each model, to assess the variability of the
simulated results. 

\subsection*{Pipeline phase III --- Analysis}

The aim is to perform an unbiased comparison of model results and
experiments using appropriate  quantitative  measures. We have assembled
a set of measures for analysing both simulated maps and those from
experimental recordings. 

\paragraph{Discrete vs continuous synapses}

All models represent the map as a set of connections in a weight
matrix. Two of the models use discrete (integer-valued) weights. For
the other two models, which use continuous valued connections, some of
the measures require the weights below certain small values to be
set to zero; these thresholds are given in Table~\ref{tab:Parameters}.

\paragraph{Map precision}

This has been measured in developing mouse by dual retrograde
injections \cite{Lyngholm2013}. Two injections of red and green beads
mark two groups of neurons in SC and the label is retrogradely
transported to RGCs. The spatial segregation of the two labelled RGC
populations are then assessed \cite{Upton2007,Lyngholm2013}.
The segregation measure is defined as the fraction of RGCs where the
nearest neighbour is the same colour. For two completely segregated
projections the value is $1$; for two overlapping projections the
value is $0.5$. Here we perform equivalent virtual injections on
simulated maps to assess map precision.

\paragraph{Contour analysis}

The distribution of synaptic labelling in the retina following dye
injection in the SC is assessed using contour analysis based on kernel
density estimates. Gaussian kernels are placed around a set of
discrete labelled points to estimate the variations in density
throughout the region. The retinal space is divided into a $100 \times
100$ grid, and each labelled point has the same weight.  The
kernel density estimate at location $\mathbf{r}$ in the grid is defined
by
\begin{equation}
\hat{f}(\mathbf{r},k) = \frac{1}{N} \frac{1}{2 \pi k^2} \sum_{i=1}^N{ \exp\left( - \frac{1}{2 k^2}|\mathbf{r}-\mathbf{r}_i|^2 \right)  }
\end{equation}
where $\mathbf{r}_i$ are the locations of the $N$ labelled neurons and
the bandwidth $k$ is chosen (using \texttt{fminsearch} in MATLAB) 
to maximise the cross-validated log-likelihood
\begin{equation}
L(k) = \sum_{i=1}^N \log(\hat{f_i}(\mathbf{r_\mathrm{i}},k))
\end{equation}
where $\hat{f_i}(\mathbf{r},k)$ is the kernel density estimate with
data point $i$ excluded. The contour curves are defined so that for
example the 25\,\% contour encloses the top 25\,\% percentile of
the total labelling from the kernel density estimate
\cite{Sterratt2013}. The readout is the retinal area covered by the
respective contour curve.

\paragraph{Lattice method analysis}

The Lattice method \cite{Willshaw2014} allows the quality, orientation
and precision of point-to-point maps to be quantified. It has been
applied to maps measured by simultaneous visual field stimulation and
Fourier-based intrinsic imaging of mouse SC \cite{Cang2008}. The
method operates on pairs of matched points located in visual field and
SC. In the experiments, the 62,500 pixels of the image determine the
point locations in the SC. For each pixel, the matching point in the
visual field is the one where stimulation excites the pixel maximally.

In the first step of the method when applied to experimental data, approximately
150 visual field points are chosen to be centres. These
are spaced approximately equidistantly, the separation being
limited by the resolution of the Fourier method. Associated with
each centre point is the group of points lying within a small circle
around it. The radius of the circle is chosen as half the separation
between nearest neighbours to ensure maximum coverage of the visual field whilst
keeping the overlap between circles small. The 150 corresponding nodes
in the SC are determined by the centroids of the projection patterns
from the points surrounding each field centre. Delaunay triangulation
is then used to construct a lattice on the field nodes, and the edges
of this lattice are then projected into the SC. Edges that cross in
the map in the SC indicate local map distortions. Connected nodes
are then removed one by one to form the largest ordered
SC submap in which no edges cross. The numbers of nodes and edges
remaining within the largest ordered submap are indicators of the
overall map quality. To give an overall measure of the orientation of
the SC map relative to the field, the mean difference in orientations of
corresponding edges in the field and the SC is computed. 

To apply the Lattice method to mappings from simulations, we take the
points in the retina to be the set of 2,000 RGC locations
$\mathbf{r}_i$. For each RGC $i$, the corresponding SC neuron $j$,
located at $\mathbf{s}_j$, is the one with the strongest connection
strength $W_{ij}$ from $i$. The Lattice method is then applied to this
set of paired points, but with 100 rather than 150 centre nodes, and
with the radius of circles in the retina being 7\,\% of the retinal
diameter. This reduction in node number is necessary due to the
smaller number of points in the simulations (2,000) than the
experiments (62,500); even so there is some overlap of the points
within the circles of neighbouring centres for the modelled data. Over
different simulations the average number of times that a single point
was used varied between 1.7 and 2.2.

To assess the global order along the anteroposterior axis we compute
the anteroposterior polarity, which is defined as the percentage
of neighbouring node pairs in the lattice that are in the correct
anteroposterior order relative to each other, given their positions on
the nasotemporal axis. The equivalent mediolateral polarity is also
calculated.

\paragraph{Visualising projections and collapse points}

In one of the experimental genotypes modelled, anterograde injections
in nasal retina resulted in two separated termination zones, while a
temporal injection gave one termination zone, see Figure~4B,H in
\citetext{Brown2000}. These authors plotted the locations of anterograde
injections of dye along the nasotemporal retinal axis against the
locations of the termination zones along the anteroposterior axis of
the SC. In \citetext{Reber2004} this experimental paradigm was
extended.  The position where the two maps converge into one was
termed the collapse point. We have automated the collapse point
detection. The nasotemporal axis is divided into 50 equal-sized bins,
and the projections originating from each bin are clustered separately
based on their termination points using the k-means algorithm. If the
distance between the means of the two clusters in the SC is larger
than 1.5 standard deviations and the smaller of the clusters contains
at least 5\,\% of the neurons, then the two clusters are considered
distinct. The algorithm defines the nasal-most bin with only one
cluster as the collapse point. In some of the cases studied there was
no collapse point present.


\subsection*{Models}

Here we describe the mechanisms of each model, listing its key
features and how they were adapted for this study.  We describe all
models in the same notation, which in some cases required a change in
notation from the published version.

\subsection*{Gierer model}

The Gierer model \cite{Gierer1983,Sterratt2013b} is a relatively
simple model of map formation that was originally formulated in 1D and
incorporated both gradients and countergradients, which are used to
define a potential that guides where synapses are placed. The model has
been extended to 2D with the countergradients removed. The competition
term also has an added decay term to prevent it from growing without
bound \cite{Nissenbaum2010}.

Each RGC axon has $N_\mathrm{term}=16$ terminals. One epoch,
equivalent to advancing time by one step, consists of examining each
terminal in the system and deciding whether to move it.
Each terminal is considered sequentially in random order.
For a terminal that connects retinal neuron $i$, with retinal
coordinates $\mathbf{r}_i$, to SC neuron $j$, with SC coordinates
$\mathbf{s}_j$ the terminal can move to one of the neighbours $j'$ of
$j$ (neighbours defined by the Delaunay triangulation on the
$N_\mathrm{SC}$ SC neurons) which has lowest potential.  The potential
at location $j’$ is
\begin{equation}
p(\mathbf{r}_i,\mathbf{s}_{j'}) = g(\mathbf{r}_i,\mathbf{s}_{j'}) +
c(\mathbf{s}_{j'})
\end{equation}
where $g(\cdot, \cdot)$ is the gradient information defined below and
$c(\mathbf{s}_{j'})$ is the level of competition at point
$\mathbf{s}_j'$ in the SC. Designating cell $j^\ast$ as the neighbour with
the lowest potential, the terminal moves to cell $j^\ast$ if this
potential is lower than the potential at its original position $j$
(i.e. $p(\mathbf{r}_i,\mathbf{s}_{j^\ast}) <
p(\mathbf{r}_i,\mathbf{s}_j)$).

\paragraph*{Gradient term}

\begin{equation}
g(\mathbf{r}_i, \mathbf{s}_j) =
R_\mathrm{A}(\mathbf{r}_i) L_\mathrm{A}(\mathbf{s}_j)
- R_\mathrm{B}(\mathbf{r}_i) L_\mathrm{B}(\mathbf{s}_j) 
\end{equation}
Here $R_\mathrm{A}$ and $R_\mathrm{B}$ are the retinal EphA and EphB
receptor concentrations, and $L_\mathrm{A}$ and $L_\mathrm{B}$ are the
SC ephrin-A and ephrin-B concentrations. The difference in signs of
the two terms indicates that A is a repulsive system, whereas B is an
attractive system.

\paragraph*{Competition term}

Competition was introduced by incorporating the term $c(\mathbf{s}_j)$
which tracks $\rho(\mathbf{s}_j)$, the density of terminals contacting
on SC neuron $j$ \cite{Gierer1983}. This term ensures an even
distribution of connections over the SC. However, this assumes
infinite memory, with the value of $c$ increasing without
bound. Following recent analysis \cite{Sterratt2013b} the term $\eta
c(\mathbf{s}_j)$ was added to weaken competition by removing the
infinite memory
\begin{equation}
\frac{\partial c(\mathbf{s}_j)}{\partial t} = \epsilon \rho(\mathbf{s}_j) - \eta c(\mathbf{s}_j)
\end{equation}

To check for a steady-state in the network, we compared the values of
$c$ with their theoretical steady-state, $c(\mathbf{s}_j) =
(\epsilon/\eta) \rho(\mathbf{s}_j)$. Simulations verified that the
maps had converged after 10,000 epochs.

There are three key parameters in the model.  $N_\mathrm{term}$ was
fixed at 16, following \citetext{Gierer1983}.  A small value of the
compensation factor $\epsilon$ was chosen to ensure that competition
is gradually enforced.  The value of $\eta$ was then chosen so that
the competition term is relatively weak in the \gMathV condition. Its
effect is ten times stronger in wild type, as \gMathV has
10\,\% of RGCs compared to wild type.

\paragraph{Summary of modifications}

Our implementation of the Gierer model has bounded competition, and
no countergradients.


\subsection*{Koulakov model}

The Koulakov model \cite{Triplett2011} uses gradient information,
competition and correlated retinal activity to define a system energy
for the current set of connections. The system evolves by repeatedly
modifying connections, favouring modifications that reduce energy. In
the Koulakov model the energy of the system consists of three terms,
representing the interaction of the chemical cues, the effect of
correlated neural activity and the effect of competition for
resources
\begin{equation}
\noindent E = E_\textrm{chem} + E_\textrm{act} + E_\textrm{comp}
\end{equation}

The chemical energy represents the repulsive interaction between EphA
and ephrin-A and the attractive interaction between EphB and
ephrin-B.
\begin{equation}
\noindent E_\mathrm{chem} = \sum_{i \in \textrm{synapses}} \left(\alpha  R_\mathrm{A}(\mathbf{r}_{\mu_i}) L_\mathrm{A}(\mathbf{s}_{\mu_i}) - \beta
 R_\mathrm{B}(\mathbf{r}_{\nu_i})  L_\mathrm{B}(\mathbf{s}_{\nu_i})\right)
 \end{equation}
 where $\alpha$ and $\beta$ define the relative strengths of the A and
 B system, $R_\mathrm{A}$ and $R_\mathrm{B}$ are the receptor
 concentrations for RGC at $\mathbf{r}$, and $L_\mathrm{A}$ and
 $L_\mathrm{B}$ are the ligand concentrations for SC neuron at
 $\mathbf{s}$; $\mu_i$,
 $\nu_i$ map synapse $i$ onto its corresponding RGC and SC neuron
 index.

 The neural activity term represents the influence of correlated
 activity on the synapses \cite{Tsigankov2006}
\begin{equation}
\noindent E_\textrm{act} = - \frac{ \gamma } { 2} \sum_{i,j \in \textrm{synapses}} C(\mathbf{r}_{\mu_i}, \mathbf{r}_{\mu_j}) U(\mathbf{s}_{\nu_i}, \mathbf{s}_{\nu_j})
\end{equation}
where $C$ represents the retinal correlation, and $U$ the pair-wise
interaction in the SC
\begin{equation}
\noindent C(\mathbf{r}_{\mu_i}, \mathbf{r}_{\mu_j}) = \exp(-|\mathbf{r}_{\mu_i} - \mathbf{r}_{\mu_j}|/b)
\end{equation}
\begin{equation}
\noindent U(\mathbf{s}_{\nu_i}, \mathbf{s}_{\nu_j}) = \exp(-(\mathbf{s}_{\nu_i}-\mathbf{s}_{\nu_j})^2 / 2a^2)
\end{equation}
where $b$ and $a$ specify the space constants.
The competition term provides an initial drive to add synapses, but
also limits the total number of synapses in the system. It is defined
as
\begin{equation}
\noindent E_\textrm{comp} = \sum_{i \in \textrm{RGCs}} \left(-500
  n_{R,i}^{0.5} + n_{R,i}^2\right) + \sum_{j\in \textrm{SC cells}} n_{SC,j}^2
\end{equation}
where $n_{R,i}$ and $n_{SC,j}$ are the number of synapses made by RGC
$i$ and SC neuron $j$. Here $i$ is summed over all RGCs and $j$
over all SC neurons. The model starts without any synapses. With a
small number of synapses initially $E_\mathrm{comp}$ is negative (term
1), reducing the total energy and favouring connection formation. As
the number of synapses increases $E_\mathrm{comp}$ grows positive
making connection formation more difficult.

Each iteration of the model has two steps. First the algorithm
attempts to add a connection between a randomly chosen pair of RGC and
SC neurons. In the second step the algorithm tries to remove a
randomly chosen existing connection. In both cases a change is
accepted with probability $P = 1/(1 + \exp(4 \Delta E))$, where
$\Delta E$ is the change in energy associated with adding or removing
the synapse. This means that changes that increase the energy are
unlikely to be accepted.

\paragraph{Summary of modifications}

The original model parameters have been adjusted to better account for
the \gIslEkp phenotype: the chemical strength was multiplied by a
factor of 4.5, and the neural activity divided by a factor of 4 (Table~\ref{tab:Parameters}). The activity term was then multiplied
by a factor of 25 to compensate for the reduced number of synaptic
pairs when the number of neurons are reduced to 2,000 from 10,000, the
value used in \citetext{Triplett2011}.

Model convergence is assessed by tracking the average spread of
postsynaptic connections in the SC for the RGC axons, or by tracking
the fraction of rejected actions, which grows as the model gets closer
to convergence. To ensure convergence each simulation was run for
10,000 epochs. The number of iterations in an epoch is equal to the
number of neurons in the simulated retina or SC, so that on average
each neuron will have had an addition and a removal step. The
total number of iterations in the system is thus $2,000 \times
10,000$.


\subsection*{Whitelaw model}

The Whitelaw model \cite{Whitelaw1981} uses chemical cues and explicit
retinal activity patterns to adapt synaptic weights in a Hebbian
fashion. The strength of the connection between RGC $i$ with location
$\mathbf{r}_i$ and SC neuron $j$ (location $\mathbf{s}_j$) is
represented by $W_{ij}$. The system starts fully connected with all weights initialised to $1$.

Chemospecificity is introduced through adhesive coefficients $M_{ij}$
between RGC $i$ and SC neuron $j$. Mimicking the expression for
chemospecificity in the original model, we define $M_{ij}$ as
\begin{equation}
M_{ij}=R_{A}(\mathbf{r}_{i})\left[\max_k(L_\mathrm{A}(\mathbf{s}_{k})) -
L_\mathrm{A}(\mathbf{s}_{j})\right]
+R_\mathrm{B}(\mathbf{r}_{i})L_\mathrm{B}(\mathbf{s}_{j})
\label{eq:WCChem}
\end{equation} 
Compared to the original formulation \cite{Whitelaw1981}, the
contribution of the A system has been altered to make it repulsive and
to ensure that the adhesive coefficients remain positive, which is
needed for synaptic plasticity (equation~\ref{eq:update}). The
B system is attractive, as was assumed for the markers in the
original 1D system \cite{Whitelaw1981}.

For each RGC $i$ and SC neuron $j$ the set
$\textrm{neigh}(\mathbf{r}_{i})$ contains the indices of RGCs
falling within a radius $r_\mathrm{R}$ of $\mathbf{r}_{i}$ (including
$i$ itself). The set $\textrm{neigh}(\mathrm{\mathbf{s}}_{j})$ was defined
similarly with a radius $r_\mathrm{SC}$ on the SC.

The algorithm proceeds on an epoch basis. For $q \in 1, ... ,N_{R}$,
RGC $q$ is the centre of activity and retinal activities, $x_i$,
are set using
\begin{equation}
x_{i} =  \begin{cases}
u & i \, \in \, \textrm{neigh}(\mathbf{r}_{q}), \mbox{where}  ~ 
u=2/{|\textrm{neigh}(r_{q})|}  \\
0 & \mbox{otherwise} 
\end{cases}
\end{equation}
This normalises the sum of RGC activity to 2, removing small spatial
variations in the density of neurons that otherwise affect
topography. This reflects the formulation in the original model where
the induced activity in the SC was scaled to be smaller than the
activity input in the retina \cite{Whitelaw1981}.

The induced activity in SC neuron $j$ is denoted by $y^{I}_j$
\begin{equation}
y_j^{I}=\sum_{i=1}^{N_r}W_{ij} x_{i}
\end{equation}

Each SC neuron receives lateral input from other SC neurons within a
radius $r_\mathrm{SC}$.
\begin{equation}
y_{j}=\frac{k}{|\mathrm{neigh}(\mathbf{s}_{j})|} \sum_{p \, \in \, \mathrm{neigh}(\mathbf{s}_{j})} y_{p}^{I}
\end{equation}
where $k$ is a proportionality constant retained from the original
model.

The Hebbian change in the weight matrix $W_{ij}$ resulting from RGC
$q$ being the centre of activity is given by
\begin{equation}
\Delta W_{ij}^{q}= \Delta t \left( \left(M_{ij}+1\right)x_{i}y_{j}-\mu y_{j}\right) 
\label{eq:update}
\end{equation}
where $\Delta t$ is the time step per activation in the retina,
$M_{ij}$ is the chemospecific adhesion (equation~\ref{eq:WCChem}) and
$\mu$ is the rate at which synapses decay due to asynchronous
activity. The addition of $1$ to $M_{ij}$ reflects the original
model's baseline gradient value which aims to ensure that when RGC $i$
and SC cell $j$ are co-active the change to the synapse strength is
positive.

The total change in $W_{ij}$ over an epoch is $\Delta W_{ij} =
\sum_{q}\Delta W_{ij}^{q}$. At the end of an epoch, $W_{ij}$ is
updated
\begin{equation}
W_{ij}(t+1)=W_{ij}(t) + \Delta W_{ij}.
\end{equation}
Any elements in $W_{ij}$ below a small threshold value
$w_\mathrm{min}$ were set to zero.  Competition is introduced to
prevent unbounded growth by normalising the matrix $W$ at the end of
each epoch. The normalisation is first done for each SC neuron, and
then for each RGC

\begin{equation}
W_{ij} \leftarrow \frac{N_\mathrm{R} W_{ij}}{\sum_i W_{ij}} \quad, \quad
W_{ij} \leftarrow \frac{N_\mathrm{SC}W_{ij}}{\sum_jW_{ij}}
\end{equation}
This order of normalisation is crucial for the formation of a double
map in the \gIslEkk phenotype: normalising first along inputs to SC
neurons maintains the effect of the different levels of EphA (which
affect the growth rate of connections) in the input RGCs. Reversing
the order of normalisation would remove the effect of the knock-in.

\paragraph{Summary of modifications}

The \citetext{Whitelaw1981} model was extended from 1D to 2D. The
chemospecificity term now contains one attractive and one repulsive
gradient.  The retinal waves were changed to activate neurons within a
radius $r_\mathrm{R}$ and the total retinal activity normalised
to maintain a constant level of activation for each wave. The weights
were normalised after each epoch instead of after each activation. The
number of neurons was increased from 20 to 2,000. The model parameters
were optimised to fit the \gIslEkk data, which requires that
postsynaptic normalisation is done before presynaptic normalisation.


\subsection*{Willshaw model}

The key concept in the \citetext{Willshaw2006} model is that SC
gradients are not fixed, but are ``induced'' by ingrowing retinal
fibres. Each RGC $i$ bears fixed quantities of EphA and EphB depending
on retinal position according to the standard gradients.  Levels of
induced marker $I^\mathrm{A}_j$, $I^\mathrm{B}_j$ in SC neuron $j$
depend on the densities of receptor in the terminals of the axons
impinging on it, weighted by the appropriate synaptic strengths
\begin{equation}
  I^\mathrm{A}_j = \sum_k W_{kj}R_\mathrm{A}(\mathbf{r}_k)/\sum_k W_{kj}
  \quad,\quad 
  I^\mathrm{B}_j = \sum_k W_{kj}R_\mathrm{B}(\mathbf{r}_k)/\sum_k W_{kj}
\end{equation}
The markers $T^\mathrm{A}_{j}$ and $T^\mathrm{B}_{j}$ represent the
densities of the ligands ephrin-A and ephrin-B in each SC neuron. Unlike
$L_\mathrm{A}$ and $L_\mathrm{B}$ in the other models, $T^\mathrm{A}$ and $T^\mathrm{B}$ vary over time,
and are produced at a rate which depends on the relationship of the
induced marker and the ligand
\begin{equation}
  \Delta T^\mathrm{A}_j = (\sigma(1 - \zeta I^\mathrm{A}_j T^\mathrm{A}_j) +
  \delta\nabla^2T^\mathrm{A}_j)\Delta t 
  \quad,\quad
  \Delta T^\mathrm{B}_j = (\sigma(I^\mathrm{B}_j - T^\mathrm{B}_j) + \delta\nabla^2T^\mathrm{B}_j)\Delta t
\end{equation}
where $\sigma$, $\zeta$ and $\delta$ are parameters and $\Delta t$ is
the time step (set equal to 1 in \citetext{Willshaw2006}). The parameter
$\zeta$ is the sole modification to the model. It is set to 3.5 to
compensate for the different magnitude of the wild type EphA gradients
in the pipeline (maximum of 1) compared to the original model (maximum
of circa 3.5).
The Laplacian operator $\nabla^2$ enforces spatial continuity through
short range interchange of markers between neuron $j$ and its
neighbours, which are defined by the links in a Delaunay triangulation
of the SC neuron locations, where edges making angles smaller than
10$^\circ$ have been removed. Each synaptic connection is updated
according to the similarity $\Phi_{ij}$ between the axonal and SC
neuron markers, and a presynaptic competitive normalisation
\begin{align}
  \Phi_{ij} &= \exp\left(-\left[\left(\zeta R_\mathrm{A}(\mathbf{r}_i) T^\mathrm{A}_j -1\right)^2 +
      (R_\mathrm{B}(\mathbf{r}_i)-T^\mathrm{B}_j)^2\right]/2\kappa^2\right)
  \\
  \Delta W_{ij} &=  \left(W_{ij} + \theta\Delta
    t\Phi_{ij}\right)/\sum_k\left(W_{ik}+\theta\Delta t\Phi_{ik}\right) - W_{ij}
\end{align}
The parameter values used (Table~\ref{tab:Parameters}) are the
same as those used in Figure~7 of \citetext{Willshaw2006}, apart from
$\zeta$.  Examination of Willshaw's code showed that $\kappa=0.0504$
rather than the $\kappa=0.72$ reported.  Simulations were run with
$\Delta t=0.1$ for $48,000$ steps. Some long simulations
($1,200,000$ steps) were also run to investigate the stability
of the maps. To set up the polarity the model requires either a weak
bias in the initial weights, or a weak bias in the gradients; here the
latter was used and the initial connection weights were sampled from
the uniform distribution. To initialise the simulation, each synaptic
strength $W_{ij}$ is drawn independently from a uniform distribution
between 0 and $10^{-4}$. The initial SC ephrins are taken from the
standard gradients,
i.e.~$T^\mathrm{A}_j=L_\mathrm{A}(\mathbf{s}_j)$. These gradients are
of a similar magnitude to those used in the original model
\cite{Willshaw2006}, though with no noise.

\paragraph{Summary of modifications}

The initial gradients used here were stronger and noise-free, compared
to before.

\clearpage
\section*{\uppercase{Results}}

\begin{figure}[!ht]
\begin{center}
\includegraphics[width=10cm]{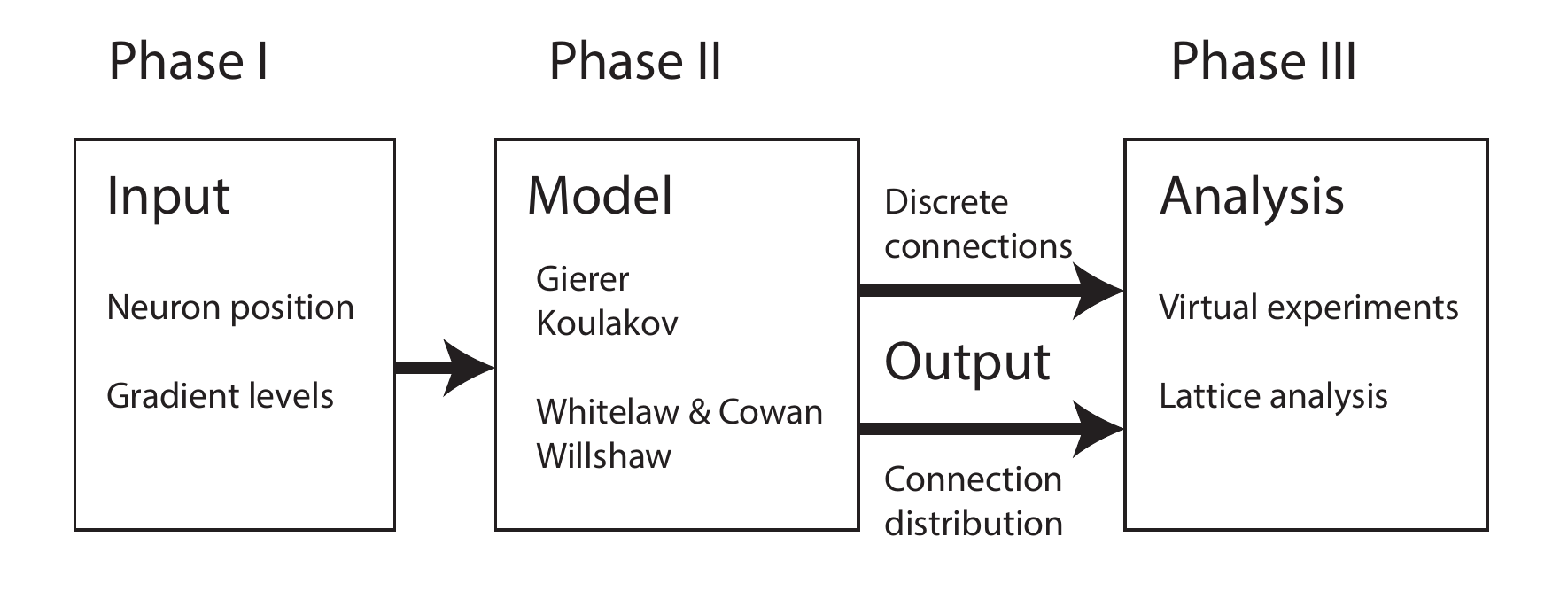}
\end{center}
\caption{ {\bf Schematic of model comparison framework.}  The pipeline
  generates neuron positions and gradient information for the retina
  and the SC which are passed to the models. After simulations finish,
  the resulting maps, irrespective of the model, are analysed using
  the same code and compared to results from experimental maps.
  Wherever possible, the map analysis method is identical for simulated
  and experimental maps.}
\label{fig:framework}
\end{figure}

\begin{figure}[!ht]
\begin{center}
\includegraphics[width=10cm]{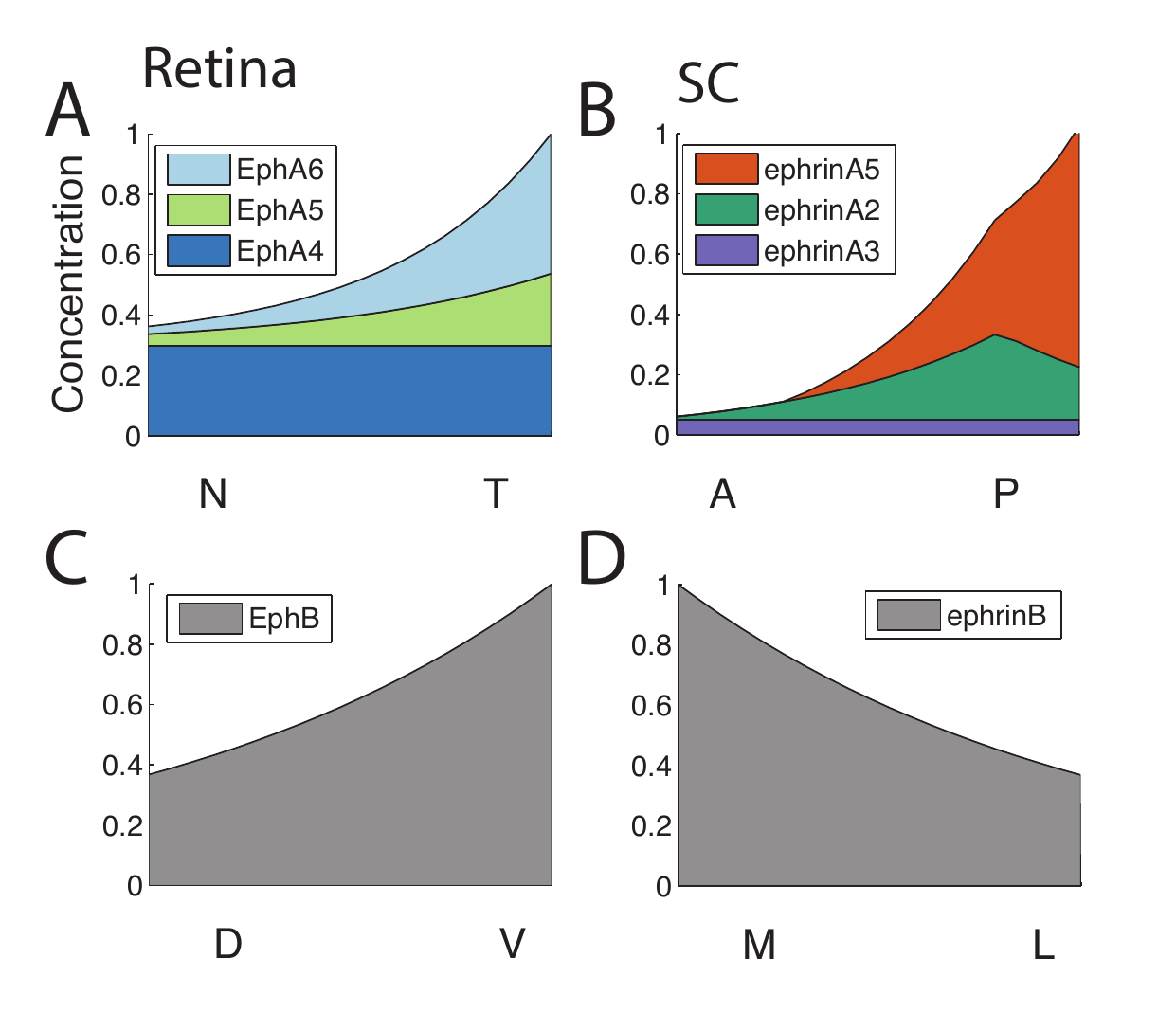}
\end{center}
\caption{ {\bf Wild type Eph and ephrin gradients used in the model
    comparison.} (A) Retinal EphA gradients, (B) SC ephrin-A
  gradients, (C) Retinal EphB gradients, (D) SC ephrin-B
  gradients. Parameters for the gradients are given in
  Table~\ref{tab:gradients}.}
\label{fig:gradients}
\end{figure}

By implementing four models and integrating them into our model
evaluation pipeline (Figure~\ref{fig:framework}, described in detail
in Methods), we can compare quantitatively each model's ability to
account for each phenotype.  The models receive similar initial
conditions for neuronal position and gradients (Figure
\ref{fig:gradients}), while the pattern of initial connectivity is set
according to each model. The resulting connectivity maps are evaluated
using the same criteria for all models, thus ensuring a fair
comparison. The four models are the Gierer model \cite{Gierer1983,Sterratt2013b},
the Koulakov model \cite{Triplett2011}, the Whitelaw model
\cite{Whitelaw1981} and the Willshaw model \cite{Willshaw2006}. For a
detailed description of each model and why it was chosen, see
Methods.

\subsection*{Wild type}

\begin{figure}[!ht]
\begin{center}
\includegraphics[width=18cm]{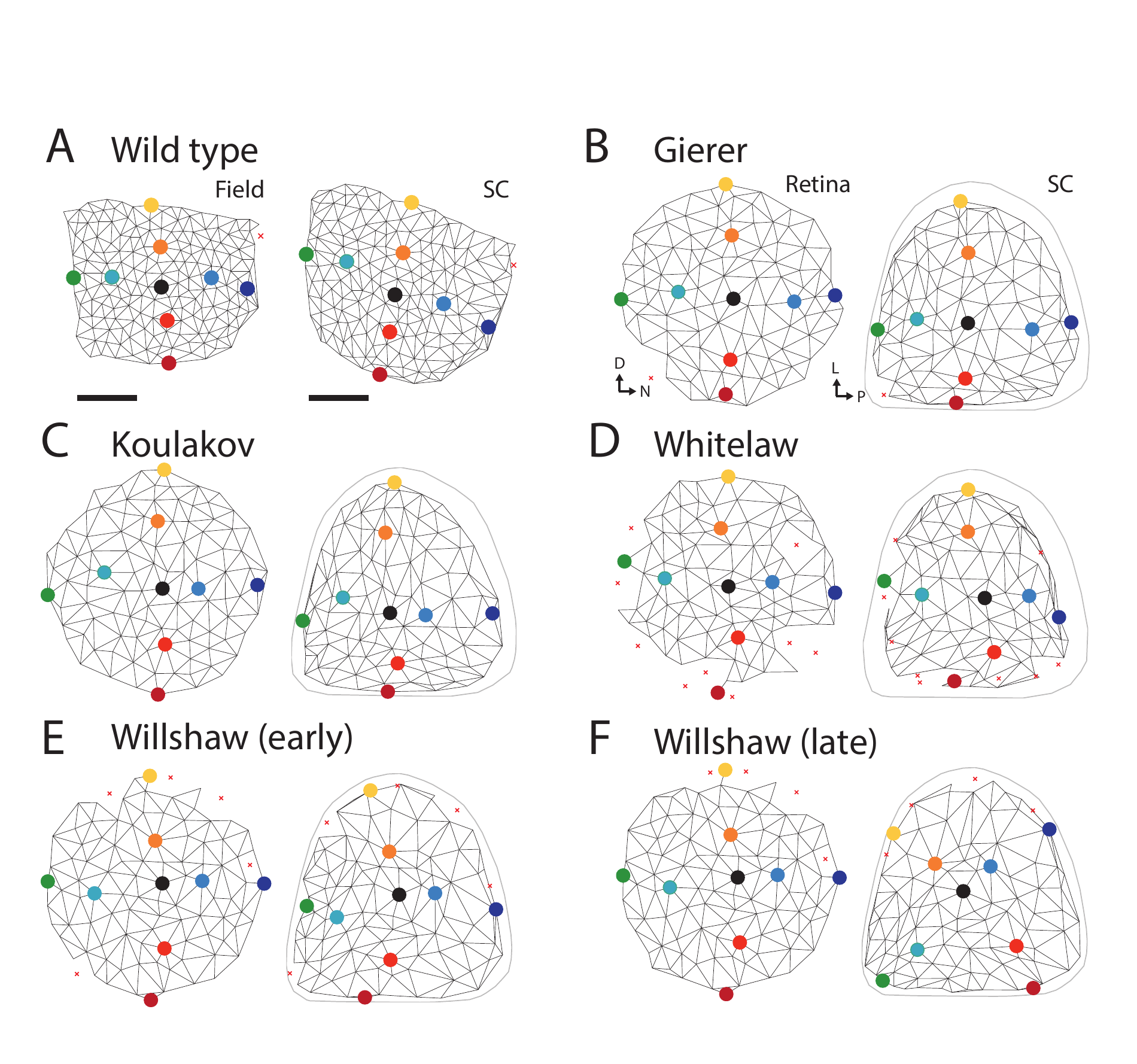}
\end{center}
\caption{ {\bf Lattice analysis of wild type map reveals a topographic
    projection.} A lattice superposed over the retina (or the field)
  is deformed by the projection onto the SC. The projection of each
  node of the lattice is the averaged projections of nearby retinal
  neurons. Nodes are connected to their neighbours by black lines. Red
  crosses mark nodes in the Lattice analysis that were removed to
  maintain a locally ordered submap. The nine coloured filled circles
  act as visual guides. (A) The adult wild type map acquired by
  intrinsic imaging shows a topographical map from field to the
  SC. The axes for the experimental data are flipped relative to
  simulated data, since nasal field projects on temporal retina. (B-E)
  Illustrative examples of the four main models are shown. (F)
  Extended Willshaw simulation (1,200,000 steps instead of 48,000),
  showing a rotation of the map. }
\label{fig:WT-lattice}
\end{figure}

The connections from retina to SC in adult wild type mice form a
topographic map as demonstrated by both electrophysiology and
intrinsic imaging \cite{Drager1976,Cang2008}. By applying the Lattice
method \cite{Willshaw2014} to this data, which involves placing a grid
over the retina (or field) and studying the deformation of its
projection onto the colliculus, we can quantify global
topographical order \cite{Willshaw2014}. The adult wild type mouse has
a topographic map in which the largest ordered submap includes the
entire field as shown in Figure~\ref{fig:WT-lattice}. The global order
is reproduced by all models, but the Whitelaw and Willshaw models have map
defects due to edge effects (Table~\ref{tab:lattice}).  

We assumed that gradients are
aligned with the standardised axis along which gradients are normally
measured (nasotemporal, dorsoventral, anteroposterior,
mediolateral). However the experimental gradients may not align with
these axes, as visual field in the SC is rotated by about $19^\circ$
\cite{Drager1976,Willshaw2014}. The simulated maps align with the axes
except for the Willshaw model which initially produces a map in the
correct orientation (Figure~\ref{fig:WT-lattice}E), but then drifts
gradually over time (Figure~\ref{fig:WT-lattice}F). This drift occurs
because both the ephrin-A and ephrin-B gradients in the SC are
modifiable, and therefore not locked to the anteroposterior and
mediolateral axes as in the other models. The orientation stabilises
so that the gradients are oriented diagonally across the SC, thus
maximising their length. The duration of the rotation is much longer
(20 times) than the period of initial organisation, so it is
questionable whether this drifting orientation is relevant.

To assess the precision of order in the retinotopic map,
\citetext{Upton2007} developed a method by which dye is
focally-injected into the SC, and then retrogradely transported to the
retina.  Small focal labels in the retina indicate a precise mapping.
The percentage of labelled retinal area is measured using contour
analysis \cite{Sterratt2013,Lyngholm2013}.  In wild-type mice, the
percentage of labelled retina decreases during development, indicating
ongoing refinement of the map \cite{Lyngholm2013}, see also
Table~\ref{tab:contour}.  We performed virtual retrograde injections
to assess precision in the simulated maps.  We found
that the maps from the Koulakov, Whitelaw and Gierer models have
similar precision to P12 mice (Table~\ref{tab:contour}).  The Whitelaw
model however showed large variations in retinal coverage due to map
imperfections. The Willshaw model projections are more diffuse and
closer to observations in P8 mice.

\begin{figure}[!ht]
\begin{center}
\includegraphics[width=12cm]{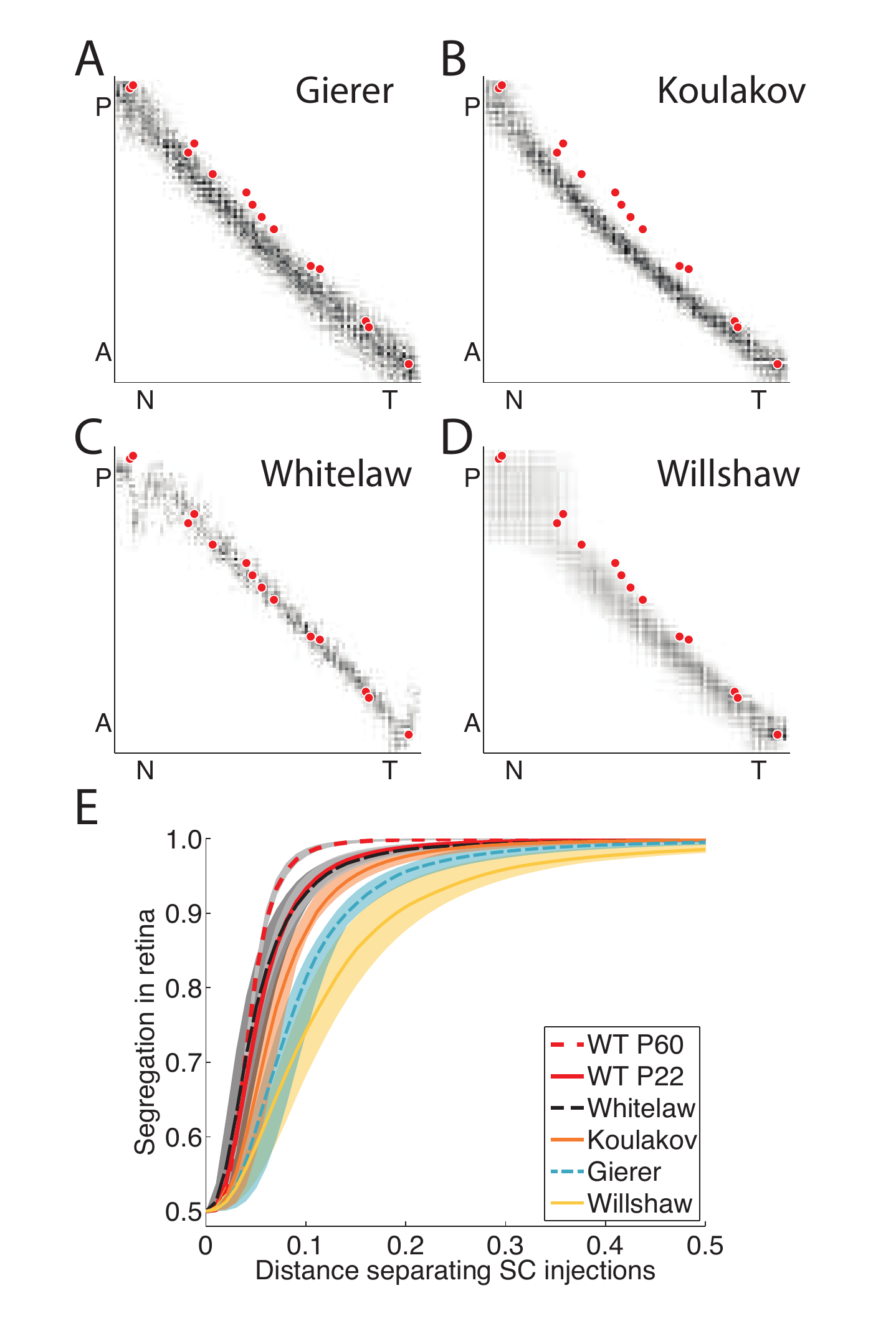}
\end{center}
\caption{ {\bf Topography and precision of the wild type map.} (A-D)
  Projection from nasotemporal axis in the retina to the
  anteroposterior axis in the SC. The black 2D histogram shows
  modelled connections; red overlaid dots are experimental data
  \protect \cite{Brown2000}. Only projections from the central third
  of the mediolateral axis of the retina are included. All models
  create a topographic map, but Gierer (A) and Koulakov (B) have a
  slight preference for anterior connections compared to the
  experimental map. Whitelaw (C) creates the most precise map, and
  Willshaw (D) the least precise map. (E) Retinal segregation of two
  retrograde injections (red and green) in the SC as a function of
  distance. The segregation measure is defined as the fraction of
  neurons whose closest neighbour has the same labelling; 0.5 means no
  segregation of the two injections, 1 means complete retinal
  segregation (See Methods). Red lines represent experimental data at
  P22 (solid) and adult P60 (dashed).  Light grey regions indicate
  confidence intervals of experimental data; ranges of simulations are
  shown in transparent colours.}
\label{fig:WT-precision}
\end{figure}

To further characterise map precision, paired dye injections are made
into SC to see how the retrogradely transported label separates in the
retina \cite{Upton2007,Lyngholm2013}.  We performed equivalent virtual
experiments: in Figure~\ref{fig:WT-precision}E the degree of
segregation between the two retinal regions is plotted as a function
of the separation of the ``virtual'' injections in the SC. The models
are designed to represent development up until eye opening at P13 in
mouse \cite{McLaughlin2003}, and no model reaches the precision
observed in P60/adult wild type mice. The Whitelaw model is most precise and
lies within the experimental range of what is seen at P22, followed by
Koulakov, then Gierer and the Willshaw model.

The difference in map precision can also be seen in the projection on
the nasotemporal axis onto the anteroposterior axis
(Figure~\ref{fig:WT-precision}A-D). Here the Willshaw model has a
wider diagonal (more spread out projections) while the Whitelaw model
has the narrowest (Figure~\ref{fig:WT-precision}E). The Gierer and
Koulakov models deviate from the diagonal, slightly favouring anterior
connections. This is presumably due to the single repulsive gradient
which, in combination with a weaker competition, makes posterior
connections less favourable.

\subsection*{Knock-in of \textit{EphA3}}

About 40\,\% of RGCs express \gIsl in a ``salt and pepper'' fashion
across the retina.  \gEphAIII is not endogenously present in retina,
but by selectively knocking in \gEphAIII in \gIsl-expressing RGCs,
neighbouring RGCs have largely different levels of EphA
\cite{Brown2000,Reber2004}. \gIslp RGCs have a higher \textit{EphA}
expression than their \gIslm neighbours, and project more anteriorly
into SC where there is less ephrin-A. Furthermore, the amount of
knocked-in \gEphAIII can be doubled in a homozygous knock-in compared
to a heterozygous knock-in.

\begin{figure}[!ht]
\begin{center}
\includegraphics[width=18cm]{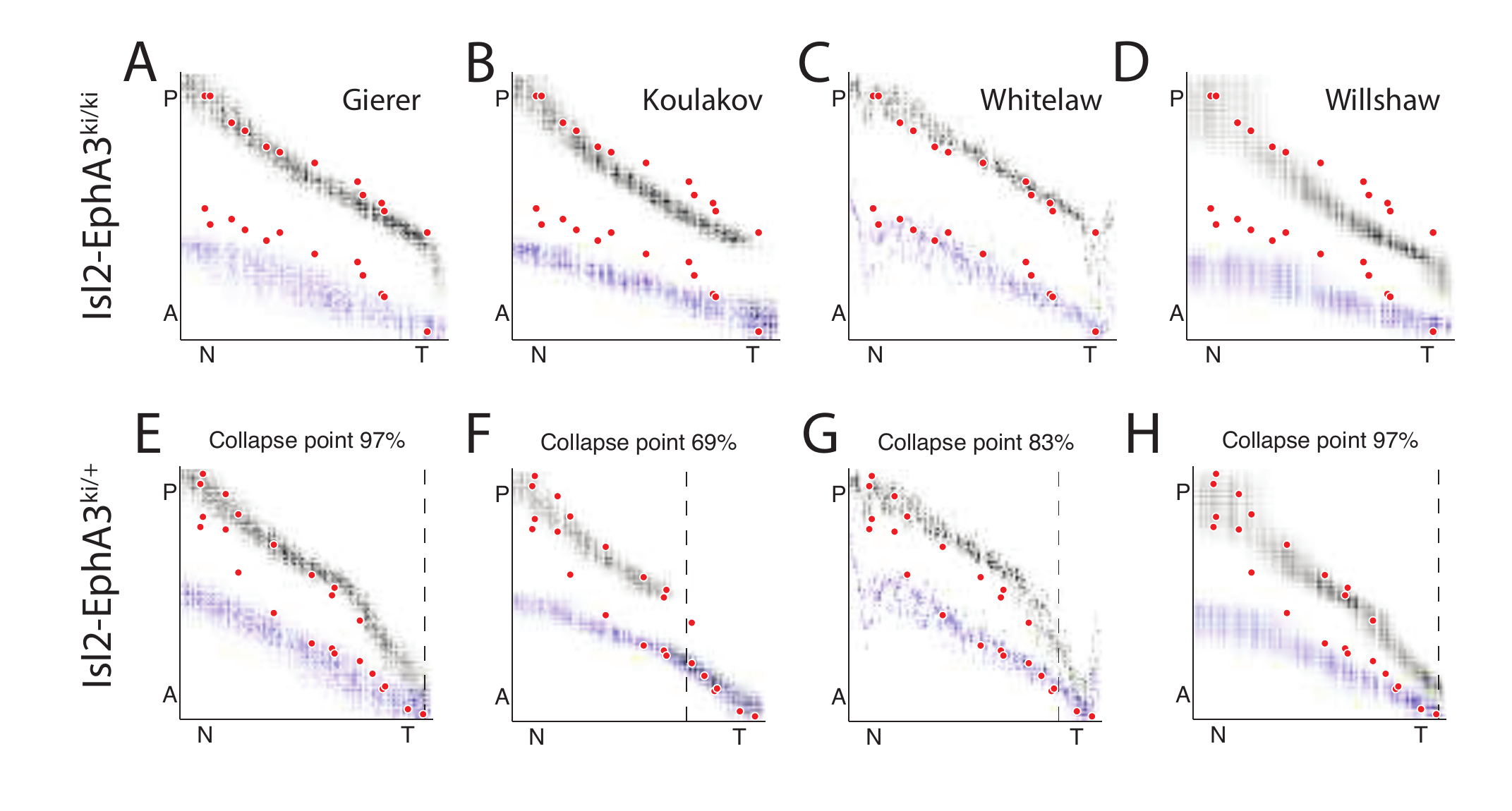}
\end{center}
\caption{ {\bf Map duplication when \textit{EphA3} is added into the
    retina.}  In the homozygous \textit{Isl2-EphA3} knock-in the
  entire map is duplicated (top row).  in the heterozygous knock-in
  (bottom row) the nasal part of the map is duplicated, but appears to
  collapse at around 76\,\% of the map.  Red dots superimposed show
  experimental data taken from Figure~5 of 
  \protect \citetext{Brown2000}. Black shows connections from \gIslm RGCs, blue
  shows \gIslp RGCs with extra \gEphAIII.  Only projections from the
  central third of the mediolateral axis of the modelled retina are
  included. The Koulakov model shows a collapse point for the
  heterozygous knock-in, the other models have a gradual merging of
  the two maps. No model has correct separation between the \gIslp and
  \gIslm maps in the SC for nasal projections. }
\label{fig:Isl2-proj}
\end{figure}

These mutants have been instructive in rejecting models based solely
on Type~I gradient mechanisms. See Choice of Models subsection in
Methods for a description of Type~I and Type~II mechanisms.  In mice
with homozygous knock-in of \gEphAIII, the map from the retina splits
into two submaps (Figure~\ref{fig:Isl2-proj}, red dots represent
experimental data). A single anterograde injection along the
nasotemporal axis in the retina generates two termination zones along
the anteroposterior axis in the SC \cite{Brown2000}. The two maps do,
however, have some overlap in the SC. A single retrograde injection
into the anterior or posterior part of SC yields one retinal
termination zone, while an injection in the central part of the SC
gives two termination zones in the retina.  Below we discuss
separately the homozygous and heterozygous knock-in of \gEphAIII.

\paragraph{Homozygous knock-in of \textit{EphA3}}

All four models generate a double map for the \gIslEkk mutant; there
are, however, subtle differences between the model results and
experimental data.  The Gierer, Koulakov and Willshaw models place the
\gIslp map (blue) anterior of the experimental data (red dots, Figure
\ref{fig:Isl2-proj}A,B,D), and the \gIslm submap appears to dip down
anteriorly at the temporal end. The Whitelaw model was optimised for
the \gIslEkk phenotype and shows a good fit to experimental data over
the majority of the nasotemporal axis (Figure~\ref{fig:Isl2-proj}C);
the exception is for extreme temporal injections which, as in the
other simulations, terminate more anteriorly.

\begin{figure}[!ht]
\begin{center}
\includegraphics[width=16cm]{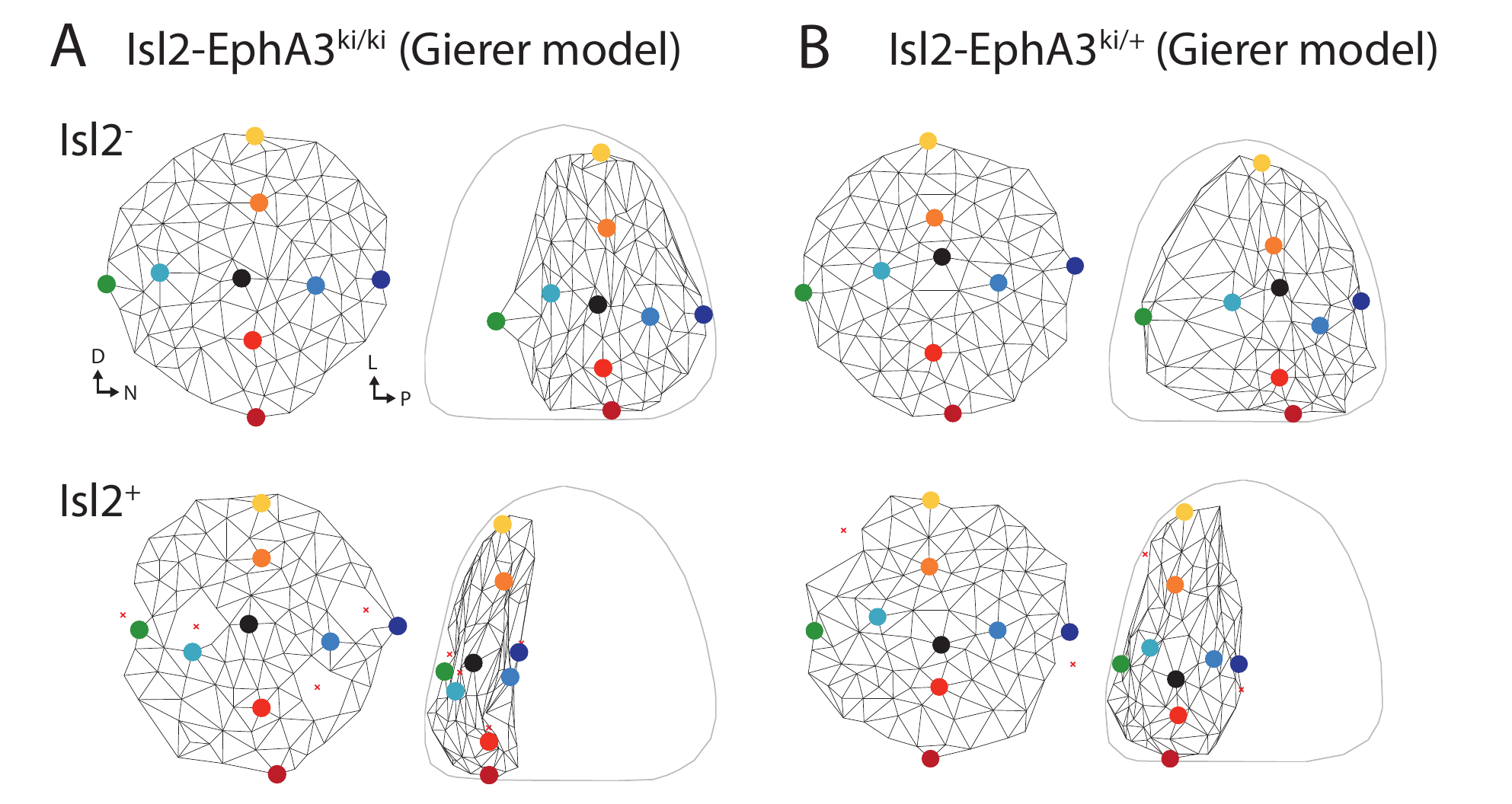}
\end{center}
\caption{ {\bf Lattice analysis of \gIslEkk and \gIslEkp.} (A) The
  extent of the \gIslp and the \gIslm submaps for the \gIslEkk
  phenotype are illustrated with the Lattice analysis.  Results are
  shown for the Gierer model and are representative of other three
  models.  (B) In the \gIslEkp the Gierer \gIslm map shows expansion
  anteriorly, and compression posteriorly.  The \gIslp map is
  restricted to the anterior end, overlapping with the \gIslm map, the
  Whitelaw and Willshaw models look similar, the Koulakov has a
  slightly lower lattice density in the middle of the \gIslm due to
  the collapse point (data not shown). See legend of
  Figure~\ref{fig:WT-lattice} for explanation of lattice plots.}
\label{fig:Isl2-lattice}
\end{figure}

The Lattice analysis shows that the \gIslp and \gIslm submaps are
almost separated for all four models
(Figure~\ref{fig:Isl2-lattice}A). In the Koulakov model this has the
consequence that a single anterograde injection gives two termination
zones, but a retrograde injection gives only a single termination
zone. The lattices show less order in the \gIslp submap for the
Whitelaw model than for the other models (Table~\ref{tab:lattice}).

\paragraph{Heterozygous knock-in of \textit{EphA3}}

In the \gIslEkp mutant there is a double map in SC which collapses
into a single map \cite{Brown2000,Reber2004} in anterior SC: nasal
retinal injections yield two termination zones, while a temporal
injection results in only one termination zone
(Figure~\ref{fig:Isl2-proj}E, red dots). The termination zones from the
nasal anterograde injections are further apart in the \gIslEkk
compared to the \gIslEkp. For retrograde injections, a single
injection in the posterior SC generates two termination zones in the retina.

All four models can reproduce the anterograde tracing experiment in
which a nasal injection yields two termination zones in the SC, and a
temporal injection gives only one termination zone. They do however
deviate from experimental results in the details. For the nasal
injection the two resulting termination zones were further apart than
in experiments. There was also a difference between how the maps
merged in the models compared to the experiments. For the Gierer,
Whitelaw and Willshaw models the two maps gradually merge (Figure
\ref{fig:Isl2-proj}E,G,H) while the Koulakov model was the only one to
exhibit a collapse point similar to what has been seen in experiments
(Figure~\ref{fig:Isl2-proj}F, $70\pm3\,\%$ along NT axis). The merge points for the
three models were located at: Gierer $95\pm3\,\%$ (7/10 simulations,
3 simulations did not merge); Whitelaw $84\pm2\,\%$; and for Willshaw
$86\pm8\,\%$ (9/10 simulations, 1 simulation did not merge). None of
the models produces two termination zones for retrograde injections in
posterior SC.

In the Koulakov model the collapse point is seen in the lattice, where
the \gIslm map is stretched (data not shown) in the
centre. For all models, the \gIslp submap does not extend
as far posteriorly as would be expected from experiments (Figure
\ref{fig:Isl2-lattice}B, showing Gierer model). The Lattice analysis
looks very similar for Whitelaw and Willshaw (data not shown) with a
stretching of the anterior part of the \gIslm submap.

\subsection*{Triple knock-out of \textit{ephrin-A}}

\begin{figure}[!ht]
\begin{center}
\includegraphics[width=18cm]{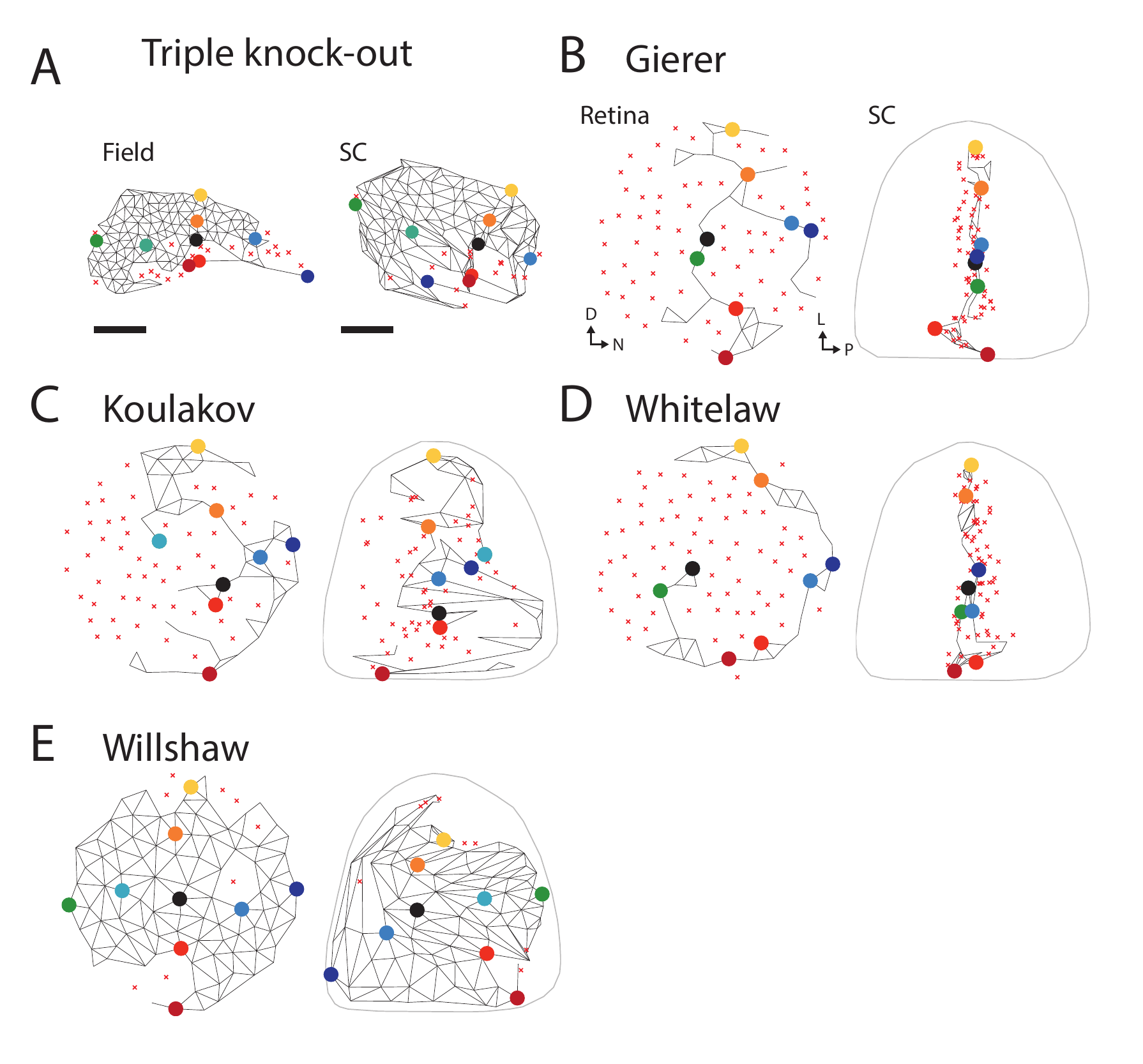}
\end{center}
\caption{ {\bf Lattice analysis of maps from TKO simulations reveals
    lack of global order.}  (A) Lattice analysis on intrinsic imaging
  data reveals order along the anteroposterior axis in the TKO
  \protect\cite{Willshaw2014}.
  (B) The Gierer model lacks order along the anteroposterior axis. 
  (C) The Koulakov model generates patches of local order, but no
  global order.
  (D) The Whitelaw model lacks order along the anteroposterior axis. 
  (E) The Willshaw model produces large patches of order but the map
  is in the incorrect orientation. See legend of Figure~\ref{fig:WT-lattice} for explanation of lattice plots.}
\label{fig:TKO-lattice}
\end{figure}

\begin{figure}[!ht]
\begin{center}
\includegraphics[width=12cm]{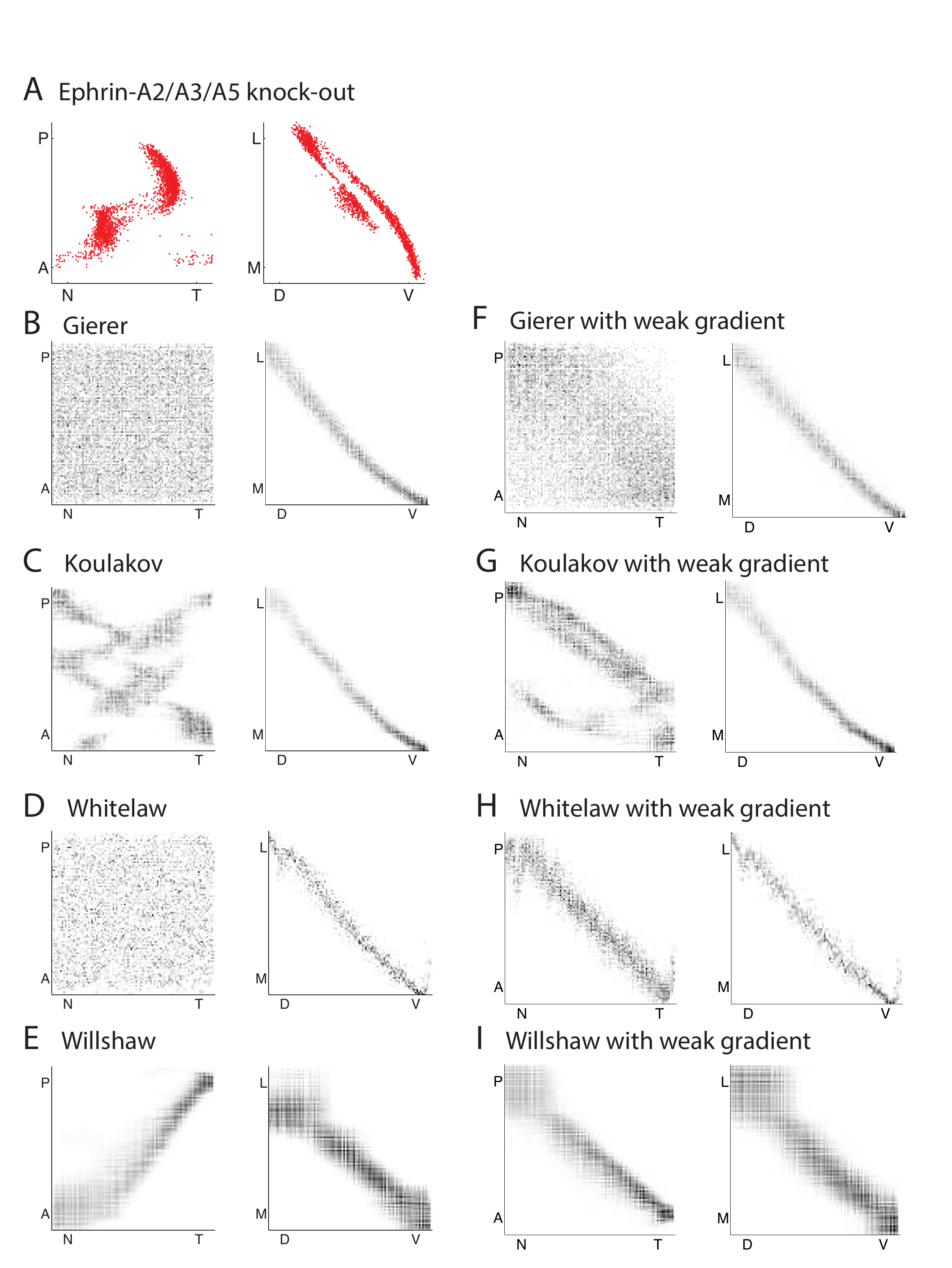}
\end{center}
\caption{ {\bf Nasotemporal and dorsoventral projections in TKO mice.}
  (A) Data from experimental intrinsic imaging \protect
  \cite{Cang2008} showing how the visual field projects onto the
  anteroposterior axis, here only the central third of the retina
  along the dorsoventral axis is used. Similarly for the visual field
  onto the mediolateral plot, where only the central third of the
  retina along the nasotemporal axis is used.  (B) The Gierer model
  maintains order along the mediolateral axis, but shows no order
  along nasotemporal axis. (C) In the Koulakov model correlated
  retinal activity joins the projections from neighbouring RGC axons
  together, creating patches of local order. (D) The Whitelaw model
  cannot produce order along the anteroposterior axis. (E) The Willshaw
  model induces gradients in the SC, forming order along the
  anteroposterior axis, but also destroying part of the order along the
  mediolateral axis in the process. In the case shown the
  anteroposterior polarity of the map is reversed. (F) The Gierer model with a weak
  anteroposterior gradient ($K = 0.01$) only has a slight increase in
  the density of projections on the diagonal. (G) The Koulakov maps
  with a weak gradient show more order, and large variations between
  runs. (H) The Whitelaw model with a weak gradient shows a complete
  diagonal. (I) The Willshaw model only needs the weak gradient to
  establish polarity and form a complete diagonal.}
\label{fig:TKO-proj}
\end{figure}

By knocking out \textit{ephrin-A2}, \textit{ephrin-A3} and
\textit{ephrin-A5} (triple knock-out, TKO) in mouse, all
\mbox{ephrin-A} ligands, which provide information about nasotemporal
mapping, are absent. The resulting map retains mediolateral order, but
initial analysis suggested very little order in the anteroposterior
direction, with patches that co-activate when one region of the retina
is stimulated \cite{Cang2008}. A more detailed analysis of the
topography using the Lattice method \cite{Willshaw2014} revealed a map
with more global order in the anteroposterior direction than initially
reported (Figure~\ref{fig:TKO-lattice}A). In these TKO
maps 10\% of the retinal positions projected to more than one
circumscribed area of colliculus, suggesting the presence of ectopic
projections \cite{Pfeiffenberger2006}.  Figure~\ref{fig:TKO-proj}A
shows the intrinsic imaging data projected onto the anteroposterior
and mediolateral axes, where the ectopic projections are apparent as
regions of the retinal nasotemporal axis which project on to multiple regions
of the anteroposterior axis.

The TKO maps from both the Gierer and Whitelaw models show no order
along the anteroposterior axis (Figure~\ref{fig:TKO-proj}B,D). The
Lattice analysis looks similar for the two models
(Figure~\ref{fig:TKO-lattice}A,C) with no regions that retain their
order when projected to the SC; instead the grid points are all
centred along the anteroposterior axis. This is also reflected in the
relatively small size of the largest ordered submaps
(Table~\ref{tab:lattice}).  The lack of order in the Gierer model
is consistent with the lack of interactions between presynaptic axons
other than competition. In the Whitelaw model an ordered map might
have been expected, since a mechanism of axon-axon interactions,
possibly mediated by neural activity, can produce ordered maps, but on
its own cannot specify global orientation \cite{Willshaw1976}. The
lack of order suggests that the specific activity mechanism
implemented in the model is weak. This is consistent with the model's
performance on the \gIslEkp mutant, where there was no collapse point.

In addition to competition, the Koulakov model also has a neural
activity term that allows for interaction between presynaptic axons,
albeit indirectly through their postsynaptic targets. The resulting
map shows patches of local order, where neural activity has joined
projections of neighbouring neurons (Figure~\ref{fig:TKO-proj}C). Some regions of the
nasotemporal axis project onto two or more distinct regions of the
anteroposterior axis, which is a hallmark of ectopic projections. The Lattice analysis
detects ordered patches (Table~\ref{tab:lattice}), and
links them together to display the largest locally ordered submap
(Figure~\ref{fig:TKO-lattice}C) but it is much smaller than
experimental submaps. There is no global order in the model maps and
the polarity of the largest ordered submap varies between different
runs.

Despite lacking global polarity cues, the Willshaw model can induce
considerable order into the largest locally ordered submap
(Figure~\ref{fig:TKO-lattice}E),
matching that seen in experiments (Table~\ref{tab:lattice}). In
addition to disrupted anteroposterior order, the
Willshaw model occasionally fails to reproduce correct mediolateral
order (Figure~\ref{fig:TKO-proj}E), which is not the case for the
other models.  Since collicular gradients adapt during simulations in
the Willshaw model, some of the order is lost in the dorsoventral axis
as the EphB and ephrin-B gradients are modified when the system tries
to induce ephrin-A gradients into an SC that initially lacks ephrin-A.

One possible explanation for the residual global anteroposterior order
in the TKO animals is that there are gradients of molecules other than
EphAs and ephrin-As along the retinal nasotemporal and collicular
anteroposterior axes which provide weak guidance information to
axons. In mouse Neuropilin 2 and Semaphorin 3F are expressed in
increasing nasotemporal and anteroposterior gradients in the retina
and SC respectively \cite{Claudpierre08}. Collapse assays show that
temporal RGC axons collapse more frequently than nasal axons in the
presence of Semaphorin 3F \cite{Claudpierre08}, though the fraction
of axons collapsing was low (4\,\% versus 12\,\%). 

If this hypothesis is true, a fairer test of the models is to
introduce a weak gradient over the rostrocaudal axis of the SC in the
homozygous TKO simulations. A simple way of simulating a weak
interaction between retina and colliculus is to replace the wild type
collicular ephrin-A gradient with a molecule having the same profile, scaled by a
factor $K<1$.

\begin{figure}[!ht]
\begin{center}
\includegraphics[width=8.5cm]{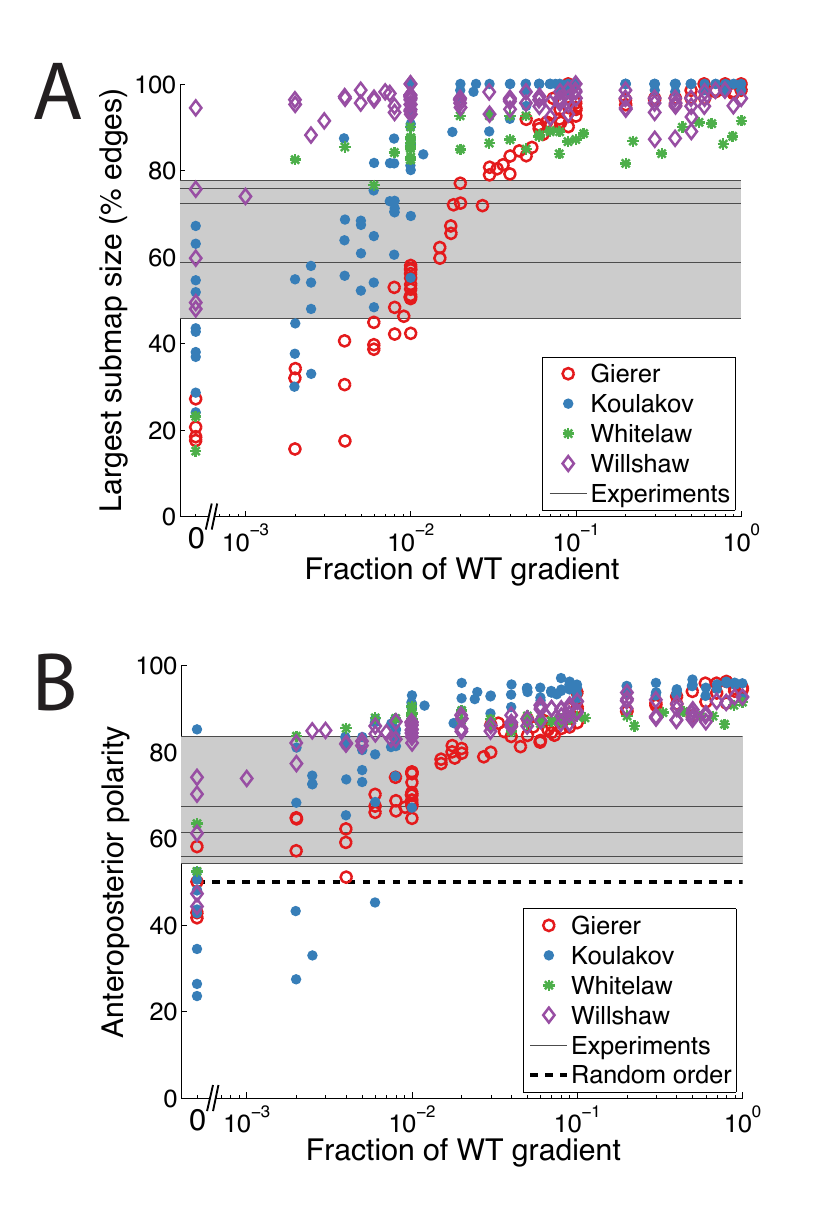}
\end{center}
\caption{ {\bf Recovering anteroposterior order in the output maps of
    the models by reintroducing a weak ephrin-A gradient into TKO.}
  (A)~Percentage of edges in the largest ordered submap as a function
  of ephrin-A reintroduced. 
  (B)~Anteroposterior order as a function of ephrin-A reintroduced. Dashed
  black line shows anteroposterior order for random maps. The grey
  region defines the range of experimental values observed. Black lines
  indicates individual experiments.}
\label{fig:TKO-fixed}
\end{figure}

Figure~\ref{fig:TKO-fixed} shows how the local order and the order
along the anteroposterior axis vary as the strength of the weak
gradient $K$ is scaled down from 1, the wild type value. Between $K=1$
and $K=0.1$ both measures remain broadly unchanged for all models.
Between $K=0.1$ and $K=0.01$, all models except for Gierer show better
quality maps than in the homozygous TKO maps. Between $K=0.01$ and
$K=0.002$, the quality of the results from Koulakov is in the range of the homozygous
TKOs and those from Gierer are worse; the other two models still
display higher quality maps. 
 
The figures show that the spread of local and global order of the
homozygous TKO maps is represented by a value of $K$ ranging from 0.03
to 0.008 for the Gierer model and 0.01 to 0.002 for the Koulakov
model. It is difficult to know how weak this gradient is relative to
wild type because lack of information about gradients and effective
interaction strengths prevents us from knowing whether a value of
$K=1$ corresponds to the wild type.

\subsection*{\gMathV knock-out}

RGC axons growing into the SC appear to compete with each other for
postsynaptic targets \cite{Gosse2008}. One way to investigate the
effect of axonal competition on map formation is to reduce the number
of innervating RGCs. In the \gMathV mutant the number of RGCs is
decreased by 90-95\,\% \cite{Triplett2011} and thus the remaining RGCs
experience reduced competition from their neighbours. The mapping in
the context of reduced competition is disrupted: instead of
innervating the entire SC, the projections are focused in the
anteromedial region \cite{Triplett2011}. It is still an open question
whether the map which forms is topographic.

\begin{figure}[!ht]
\begin{center}
\includegraphics[width=18cm]{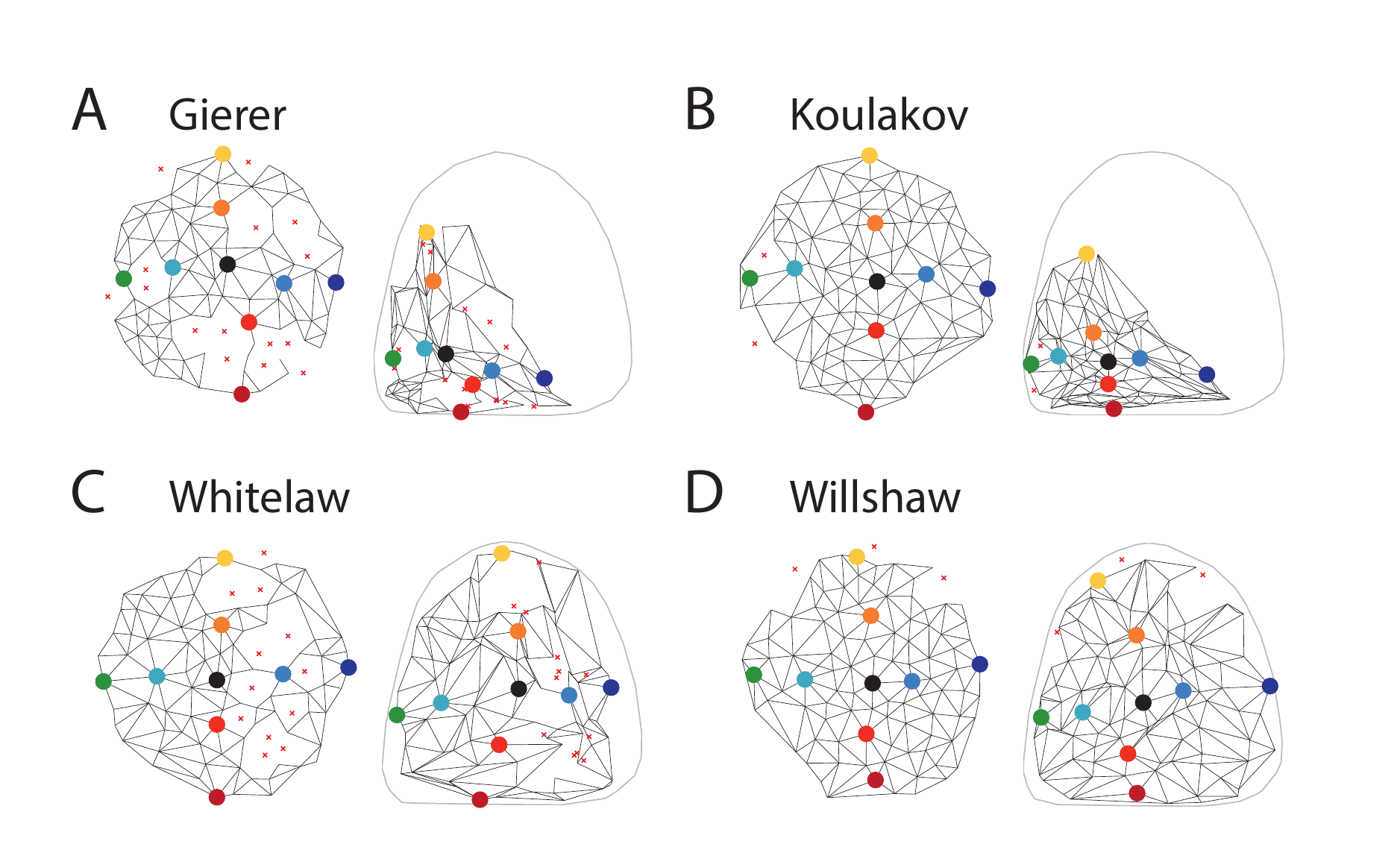}
\end{center}
\caption{ {\bf Lattice analysis of \gMathV simulations.} The Gierer
(A) and Koulakov (B) models show a anteromedial localisation of the
maps in the SC for \gMathV, with the Koulakov map being more ordered
($78.1\pm 8.4$ vs $28.4 \pm 7.6$ nodes in largest ordered
submap). Both the Whitelaw (C) and Willshaw (D) models fail to produce
the \gMathV phenotype, instead projecting across the entire SC.}
\label{fig:Math5-lattice}
\end{figure}

The Gierer model captures the anteromedial confinement of the
projection (Figure~\ref{fig:Math5-lattice}A). All RGC axons have
highest affinity anteriorly, and it is only through competition that
some of them are pushed more posteriorly. However with the reduced
population in the model, the remaining neurons can terminate more
anteriorly than they would normally do. The Koulakov model also
reproduces the anteromedial bias of the projection (Figure
\ref{fig:Math5-lattice}B). Comparing the Gierer and Koulakov maps we
see that they cover a similar fraction of the SC (99\,\% of synapses
cover $48.5\pm0.4\,\%$ vs $50.0\pm0.4\,\%$). There is however more
order in the Koulakov map than in the Gierer map (Table
\ref{tab:lattice}). The RGCs in the Whitelaw model innervate the
entire SC (Figure~\ref{fig:Math5-lattice}C) because postsynaptic
normalisation ensures that all SC neurons receive input. There is
some order retained in the largest ordered submap, but less so than in
the Koulakov model. Like the Whitelaw model the Willshaw model also
contains mechanisms that ensure that the entire SC is innervated
(coverage $98.5\pm0.4\,\%$, Figure~\ref{fig:Math5-lattice}D), and most
of it is topographically ordered (Table~\ref{tab:lattice}).



\subsection*{Summary}

In this study the Gierer, Koulakov, Whitelaw and Willshaw models of
retinotopic map formation have been evaluated quantitatively on a set
of phenotypes. In each model the same set of parameter values was used
for all simulations. Three of the four models were fitted to one of
the phenotypes: Gierer \gMathV, Koulakov \gIslEkp and Whitelaw
\gIslEkk. The Willshaw model did not require any additional
fitting. 

Our results are summarised in Table~\ref{tab:summary}. We find that
all models can account for wild type maps and the homozygous \gEphAIII
knock-in maps. The Koulakov model was the only model to generate a
collapse point in \gIslEkp maps. The Willshaw model was the only model
to produce the internal order seen in TKO maps without any extra cues,
but it does not capture the global polarity. By adding a weak gradient
(which might correspond to retinal Neuropilin 2 and collicular
Semaphorin 3F) all models could produce internal order and global
polarity. The Gierer and Koulakov models can produce the compression
of the map into the anteromedial part of SC seen in the \gMathV
phenotype.

\clearpage

\section*{\uppercase{Discussion}}

Over the 70 years during which the retinocollicular or retinotectal
projection has been used as a model system for the development of
ordered nerve connections, many computational models have been
proposed. Several reviews have synthesised properties of computational
models proposed in the last forty years to account for the development
of retinotopic maps
\cite{Swindale1996review,Goodhill2005review,Goodhill2007review}.
However, it has been difficult to assess and compare models because
either different models were formalised in incompatible ways or they
were designed for a specific set of data or key experimental data was
not available to test them. 

Therefore, before embarking on generating new models, we have aimed to
explore rigorously whether any model could account for all known data
on retinocollicular maps in mouse. To do this we have developed an
open computational framework to compare quantitatively the results
from theoretical models of retinotopic map formation against
experimental data.  We chose a set of well documented experimental
data for the mouse visuocollicular system as reference experimental
data. Exhaustive testing of all previous retinotopic map formation
models is infeasible and so we selected four representative models
that we believe collectively sample the major mechanisms hypothesised
for map formation.  In choosing models we had to eliminate those which
were not described in sufficient detail to enable us to simulate a 2D
version and those in which there was no explicit representation of
gradients. The four models chosen are: the Gierer model (1983), the
Koulakov model (2006--2011), the Whitelaw model (1981) and the Willshaw
model (1977--2006). The previously published 1D versions of both the
Gierer and the Whitelaw models required considerable extension to
enable them to reproduce 2D maps.

\subsection*{Summary of model performance}
\label{ModelComparisonDevNeuro_DW:sec:summ-model-perf}

\begin{enumerate}
\item All models could replicate wild type maps and produce
  qualitatively correct double maps seen in \gIslEkk mice.  The
  Whitelaw model produced the best match to the \gIslEkk maps,
  although its parameters were optimised for this condition.

\item The Koulakov model alone reproduced a collapse point
  in \gIslEkp mice, due to the strong activity-dependent mechanism.
  The relative contribution of activity-dependent mechanisms in the
  Whitelaw model was too weak to generate collapse points. Both the
  Gierer and Willshaw models lack a mechanism that conveys information
  about distance between pairs of retinal cells independently of
  gradients and so were not expected to reproduce the collapse point.

\item No model could account for the consistent, residual global order
  along the rostrocaudal axis in maps when all ephrin-A ligands were
  removed \cite{Cang2008,Willshaw2014}.  However, both the Koulakov
  and Willshaw models produced some order along the
  anterior--posterior axis, though its origin was quite different in
  the two cases: from correlated neuronal activity (Koulakov); from the
  spatial continuity enforced by diffusion of collicular markers
  (Willshaw). By reintroducing a weak rostrocaudal gradient back into the
  SC a largest ordered submap consistent with experiments can be
  produced by the other two models.

\item The Gierer and Koulakov models both reproduce the \gMathV
  phenotype where the projection is restricted to one portion of the
  colliculus. In the Whitelaw model the strong postsynaptic
  normalisation counteracted the effect of the Type~II mechanism to
  cluster axons at the temporal end; in the Willshaw model,
  diffusion of collicular labels caused the projection to spread,across the
  entire colliculus.
        
\end{enumerate}

\subsection*{Insights into mechanisms of mapping}

We now summarise what we have learnt about the mechanisms of map
formation and what components any new model should possess.  We do
this in terms of the five component mechanisms mentioned in the
Introduction.

\begin{description}
\item [Chemoaffinity] We found that two combinations of
    chemoaffinity account for the formation of wild type maps and the
    \gIslEkk maps.

\begin{itemize}
\item Type II affinity \cite{Prestige1975} with single set of
  gradients and a competitive mechanism (Gierer, Whitelaw, Koulakov);
  Gierer and Koulakov also gives a restricted projection in the
  \gMathV case.
\item Type I affinity with a single set of retinal gradients together with variable collicular gradients (Willshaw)
\item Models employing countergradients cannot be ruled out but those
  using fixed gradients with no plasticity are excluded by the
  \gIslEkk data.
\end{itemize}

\item [Spontaneous neural activity and Hebbian synapse formation] The
  main effect we observed in introducing a mechanism involving neural
  activity is that it enables the Koulakov model to reproduce the
  collapse point in the \gIslEkk map. Activity seems also to be
  necessary for the refinement of initial axonal arbors
  \cite{Lyngholm2013}. The representation of neural activity in both
  the Whitelaw and Koulakov models is quite abstract and so is hard
  to relate to experimental data. A more explicit representation
  (e.g.~spike times or bursting activity of neurons) would allow
  retinal wave data to be used more directly in models and allow for a
  more direct comparison with \betaIIKO mice and other
  activity-altering genotypes.

\item [Competition] All models tested incorporate a competition
  mechanism to give flexibility in the map. In the context of the
  neuromuscular system, competition mechanisms have been classified as
  consumptive competition (for neurotrophic factors) or interference
  competition, either for space, or where axons have direct negative
  interactions \cite{Ooyen2001}. By manipulating expression levels of
  the neurotrophin BDNF in individual cortical neurons it has been
  shown that BDNF helps the cells compete for inputs, and thus acts as
  a target for consumptive competition \cite{English2012}.  Since BDNF
  and TrkB are expressed in the colliculus and retina respectively
  \cite{Marler08}, there is therefore consumptive competition in the
  retinocollicular mapping. Theoretical competition rules which
  maintain the total synaptic weight assigned to all the synapses of a
  neuron at a constant level can be seen as an approximate
  implementation of consumptive competition; of the models studied the
  Whitelaw and Willshaw have this mechanism. The Koulakov model has a
  stochastic implementation of the mechanism. In contrast, the Gierer
  model has a form of competition more akin to direct negative
  interactions or space, for which we are not aware of any direct
  experimental evidence in the retinocollicular system. Manipulating
  competition rules in models could be used to check the intuition that
  reducing the expression in a portion of the SC might be expected to
  magnify the map from the retina in this region.

\item [Ordering of fibres in the optic tract] None of the models
  examined incorporates such a mechanism although in the original
  version of the Willshaw model \cite{Malsburg1977} it was proposed
  that the fibre ordering could specify the overall orientation of the
  map. Evidence for ordering across the mediolateral dimension of the
  tract \cite{Plas2005} could be used in future models. These would
  have to incorporate the three dimensions of fibre growth and
  innervation which so far has been neglected in models.

\item [Axon-axon interaction] Here we mean chemospecific signalling
  between RGC axons in the colliculus, either directly, as modelled by
  \citetext{Yates2004} and \citetext{Gebhardt2012}, or indirectly in the Willshaw model
  through the labels induced from retinal axons into the
  colliculus. In direct interactions, Eph receptors on growing axons
  are activated by ephrin ligands on nearby retinal axons and the
  strength of this effect is supposed to grow as more axons fill the
  colliculus.  Given a choice, temporal axons prefer growing on
  temporal retinal substrate, while nasal axons grow on both temporal
  and nasal retinal substrate \cite{Bonhoeffer1985} and there is also
  direct evidence for axon-axon interactions from time lapse imaging
  of interactions between growing RGCs \cite{Raper1990} as well as
  modelling arguments \cite{Weth2014}.  

  \citetext{Gebhardt2012} included direct
  axon-axon interactions in a model with gradients of retinal Eph and
  collicular ephrin and countergradients of retinal ephrin and
  collicular Eph. Without axon-axon interactions the parameters of the
  gradients and countergradients had to be matched to produce wild
  type maps. Axon-axon interactions could compensate for this,
  although this may depend on a precise matching of parameters
  \cite{Sterratt2013b}. Nevertheless, this demonstrates that axon-axon
  interactions may confer flexibility on map formation, even without
  competition. Because we did not include countergradients, direct
  chemospecific axon-axon interactions were beyond the scope of our
  study, though they can be modelled using our pipeline.

  The indirect axon-axon interactions in the Willshaw model, coupled
  with competition and a Type I affinity mechanism, gave very robust
  map formation -- more robust to knockout of \gMathV and
  \textit{ephrin-A}s than the experimental phenotypes. In the case of
  \gMathV, this robustness appears to be due to the Type~I affinity
  mechanism. Once the collicular gradients have been set up, there is
  no part of the colliculus which is preferred by all axons. In
  contrast, in models with competition and Type~II affinity, all axons
  prefer anterior colliculus; in the \gMathV knockout, competition is
  not strong enough to then force out the less-repelled nasal axons,
  as in wild types.

\end{description}

In summary, the models we examined used mechanisms of chemoaffinity,
neural activity and competition each combined into a single model
which accounted for most of the experimental data we examined, using
the same set of parameter values for all the data. The main class of
result that was not accounted for was the residual order seen in the
homozygous triple knockout map although these data could be fitted by
an additional weak gradient. This could be provided, for example, by
retinal and collicular gradients of Neuropilin 2 and Semaphorin 3F in
mouse \cite{Claudpierre08}, or possibly by Repulsive Guidance
Molecule, which in chick is expressed in a graded fashion and repels
temporal RGC axons \cite{Monnier2002}.  Another candidate is
\textit{Engrailed} which is expressed in an
anteroposterior gradient in chick tectum
\cite{Wizenmann2009,Stettler2012}. This does not exclude the
possibility that other factors, such as time of axon arrival, are
involved in generating nasotemporal map polarity.

\subsection*{Experimental considerations}

As most experimental work in topographic map formation is now
undertaken in mouse, we focused on curating the experimental data
available in the literature including wild type, \gIslEkk, \gIslEkp,
triple \gTKO knock-out and \gMathV.  We found that although there are
many other documented disruptions to the retinotopic map, often there
were few quantitative characterisations of the data, though this may
partly be due to the limitations of experimental techniques and the
variability of phenotypes. For example, a common phenotype observed in
mutant mice is that of ectopic projections
\cite{Feldheim2000,Frisen1998}. Here the raw data are images of
colliculi stained by DiI transported by axons from retinal injection
sites. Each individual has only one injection site and it would appear
that there is considerable variability between individuals, so it is
not possible to construct one composite map, as in the case of the
knock-in mutants.  To move from a qualitative to a quantitative
characterisation of ectopic projections would require significant
effort and, ideally, the availability of raw image data would allow
for various methods of determining the location of dye spots to be
tested.

The ability to obtain whole maps from individuals using functional
imaging gets around the issues of inter-individual variability, though
brings with it the problem of inferring anatomy from functional data.
Ectopic projections defined functionally have been analysed
quantitatively in TKO Fourier imaging data \cite{Willshaw2014};
applying this technique to the ephrin-A knockout data
\cite{Feldheim2000,Frisen1998} may prove fruitful.

Our modelling is dependent on (and limited
by) quantitative characterisation of the molecular gradients, notably
retinal EphA receptors \cite{Reber2004}.  Our best guesses at
parameters for the remaining Eph and ephrin gradients
(Table~\ref{tab:gradients}) can be replaced with experimental findings
once they become available.  Currently we have excluded
countergradients from our models because (a) there is limited data
about their expression levels, and (b) recent theoretical findings
suggest that competition and countergradients can be traded off
against each other \cite{Sterratt2013b}.

To investigate the role of activity in the formation of a collapse
point in \gIslEkp it might be instructive to combine this mutant with
\betaIIKO mice, where spontaneous activity is perturbed significantly
\cite{Stafford2009}.  It would be interesting to assess whether the
two maps normally seen in \gIslEkk mice converge into one, or if the
collapse point in the maps of \gIslEkp mice moves.  Unfortunately
\betaIIKO maps are inherently diffuse, so it might not be possible to
separate the two cases in the combined mutants.

Finally, one limitation of our current approach is that although the
model provides full access to the developmental time course, currently
we have limited developmental dynamics from the experimental system.
We might expect that during the critical period of map formation in
mouse, whilst the map is changing, other aspects of the system change
too.  For example, currently we assume that molecular gradients are
fixed, but these might flatten over time \cite{Rashid2005}. This could
change the balance between mechanisms driven by activity and chemical
cues.

\subsection*{Future work and challenges}
\label{ModelComparisonDevNeuro_DW:sec:future-work-chall}

There are three obvious directions in which the work can be taken:

(i) Whilst a combination of chemoaffinity, neural activity and competition
accounts for the data (within the limits stated), it may be that other
combinations also comprising mechanisms of fibre pre-ordering and/or
axon-axon interaction can also account for the data. Then it should
be possible to provide predictions to distinguish between the
different possible models.

(ii) For each of the four models we have found a set of parameter values
that can be used to produce satisfactory maps on our current data
sets. The challenge would then be to test out these models using the
same set of parameter values on new data when available.

(iii) Quantifying data from ephrin-A knockouts and challenging the
models with this data. The inter-individual variability will prove a
challenge; the question is how to match a distribution of models to a
distribution of data.

Unbiased quantitative evaluations of existing models using the
framework that we have developed will allow us to see how the
different models perform, and will help us guide future modelling
efforts. Using a curated set of experimental data makes it easier to
test a computational model and, when new experimental data becomes
available, predictions can be generated on all models.  We hope that
our open-access pipeline will inspire further unification of models to
help comparison, and increase reproducibility \cite{Stevens2013b}.

\clearpage
\bibliography{ModelComparison-arXiv}

\clearpage

\newgeometry{margin=1.5cm}
\section*{\uppercase{Figures}}


\clearpage

\clearpage

\clearpage

\clearpage


\clearpage

\clearpage

\clearpage

\clearpage


\clearpage


\clearpage

\clearpage

\begin{table}
\begin{center}
\begin{tabular}{lrrl}
\hline
Parameter & Default value & Original value & Meaning\\
\hline
\multicolumn{4}{l}{\textrm{General}}\\
$N_\mathrm{R}$ & 2,000 & n/a & Number of RGCs\\
$N_\mathrm{SC}$ & 2,000 & n/a & Number of SC neurons\\
$d_\mathrm{R}$ & 0.0139 & n/a & Exclusion distance in retina\\
$d_\mathrm{SC}$ & 0.0119 & n/a & Exclusion distance in SC
\vspace{2mm}\\
\multicolumn{4}{l}{\textrm{Gierer}}\\
$N_\mathrm{term}$ & 16 & 16 & Number of terminals made by each RGC\\
$\epsilon$ & 0.005 & 0.005 & Growth rate for competition \\
$\eta$ & 0.1 & n/a & Decay rate for competition \vspace{2mm}\\
\multicolumn{4}{l}{\textrm{Koulakov}}\\
$\alpha$ & 90 & 20 & Chemical strength of A-system\\
$\beta$ & 135 & 30 & Chemical strength of B-system\\
$\gamma$ & 25/80 & 1/20 & Strength of activity interaction\\
$b$ & 0.11 & 0.11 & Retinal correlation distance\\
$a$ & 0.03 & 0.03 & SC interaction distance \vspace{2mm}\\
\multicolumn{4}{l}{\textrm{Whitelaw}}\\
$r_\mathrm{R}$ & 0.07 & n/a & Radius of retinal activity\\ 
$r_\mathrm{SC}$ & 0.0289 & n/a & Radius of SC interaction\\ 
$\mu $ & 0.1 & 0.1 & Weight decay rate\\
$\Delta t$ & 0.0001 & [0.05, 0.5] & Integration time step \\
$w_\textrm{min}$ &  0.00001 & 0.009 & Minimum synapse strength \vspace{2mm}\\
\multicolumn{4}{l}{\textrm{Willshaw}}\\ 
$\sigma$ & 0.05   & 0.05  & Induced marker source strength \\
$\delta$  & 0.01   & 0.01  & Induced marker diffusion strength \\
$\theta$ & 0.1    & 0.1   & Speed of weight update \\
$\kappa$ & 0.0504 & 0.0504 & Sharpness of receptor-ligand comparison \\
$\zeta$   & 1 & 3.5 & Scale of induced marker and ligand interaction \\
$\Delta t$ & 1 & 0.1 & Integration time step \\
$w_\textrm{min}$ &  0.001 & n/a & Minimum synapse strength\\
\hline
\end{tabular}
\caption{Parameter values used in the models.  Column 2 denotes the
  parameter values used in this study, compared to those used in
  previous studies (column 3).
\label{tab:Parameters}
}
\end{center}
\end{table}

\clearpage

\begin{table}[h]

{\renewcommand{\arraystretch}{1.3}
\begin{tabular}{l...lp{8cm}}
\hline
\multicolumn{1}{l}{Protein} &
\multicolumn{1}{l}{$G_0$} &
\multicolumn{1}{l}{$G_1$} &
\multicolumn{1}{l}{$G_2$} &
\multicolumn{1}{l}{$G_3$} &
\multicolumn{1}{l}{Source}\\
\hline
\multicolumn{6}{l}{Retinal Eph gradients} \\
EphA4 & 1.05 & 0 & 0 & 1 &  Measured \cite{Reber2004} \\
EphA5 & 0 & 0.85 & 1.8 & 1 & Measured \cite{Reber2004} \\
EphA6 & 0 & 1.64 & 2.9 & 1 & Measured \cite{Reber2004} \\
EphA3 from \gIslEkk & 1.86 & 0 & 0 & 1 &  Measured \cite{Reber2004} \\
EphA3 from \gIslEkp &  0.93 & 0 & 0 & 1 & Measured \cite{Reber2004} \\
EphB & 0 & 1 & 1 & 1 & Postulated \cite{McLaughlin2005} \vspace{2mm} \\
\multicolumn{6}{l}{SC ephrin gradients} \\
ephrin-A2 & -0.06 & 0.35 & 2 & 0.8 & 
Estimated \cite{Frisen1998,Feldheim2000,Hansen2004} \\ 
ephrin-A3 & 0.05 & 0 & 0 & 1 & Estimated \cite{Pfeiffenberger2006,Triplett2012}\\ 
ephrin-A5 & -0.1 & 0.9 & 3 & 1 & Estimated \cite{Frisen1998,Feldheim2000,Hansen2004,Rashid2005}  \\ 
ephrin-B & 0 & 1 & 1 & 0 & Postulated \cite{McLaughlin2005} \\ 
\hline
\end{tabular}
}
\caption{Quantitative representation of Eph and ephrin
  gradients in RGCs and SC.  Retinal EphA gradients were
  measured \protect \cite{Reber2004}; `estimated' values are our
  measurements from   published figures; `postulated' means gradients
  have been   proposed based on limited data.  The gradient at a point
  $x$ is given by $G(x) = \mathrm{max}(0,G_0 + G_1 \exp(-G_2 \lvert
  x-G_3\rvert))$ where $x \in [0,1]$ is
  the position along an axis (nasotemporal,
  dorsoventral, anteroposterior or mediolateral). The gradients of
  each subtype are summed together. The summed gradients were
  normalised such that the peak value for each of the summed WT gradients were 1.
  This scaling was kept for all phenotypes. Thus for EphA3
  knock-ins the peak gradient were larger than 1, and for knock-out
  phenotypes the peak gradient was less than 1.}
\label{tab:gradients}
\end{table}

\clearpage

\begin{table}
  \centering
  \begin{tabular}{lll}
    \hline
    &
    \multicolumn{2}{c}{Largest ordered submap size} \\
    Genotype / Model & Nodes (\%) & Edges (\%) \\
    \hline
    \multicolumn{3}{l}{Wild type} \\
    Experiment & $98.3\pm2.1$& $99.5\pm2.1$\\
    Gierer & $97.8\pm3.9$ & $99.3\pm1.2$\\
    Koulakov & $99.2\pm2.5$ & $99.9\pm0.5$\\
    Whitelaw & $59.8\pm8.4$ & $88.1\pm2.9$\\
    Willshaw & $79.6\pm9.1$ & $94.8\pm2.3$\\
    & & \\
    \multicolumn{3}{l}{\gIslEkk} \\
    Experiment & -  &  - \\
    Gierer & $99.0\pm2.2$ & $99.7\pm0.7$\\
    & $60.6\pm10.3$ & $87.8\pm3.6$\\
    Koulakov & $97.6\pm3.1$ & $99.3\pm1.0$\\
    & $51.9\pm11.8$ & $81.0\pm7.0$\\
    Whitelaw & $60.1\pm6.8$ & $88.5\pm3.0$\\
    & $29.0\pm7.8$ & $73.8\pm4.2$\\
    Willshaw & $84.8\pm7.5$ & $95.8\pm2.6$\\
    & $77.5\pm9.7$ & $93.5\pm3.3$\\
    & & \\
    \multicolumn{3}{l}{\gIslEkp} \\
    Experiment & -  &  - \\
    Gierer & $94.7\pm7.0$ & $98.3\pm2.1$\\
    & $77.6\pm8.6$ & $93.6\pm2.5$\\
    Koulakov & $95.7\pm3.3$ & $98.8\pm0.8$\\
    & $79.6\pm11.9$ & $93.1\pm5.2$\\
    Whitelaw & $54.5\pm5.1$ & $87.8\pm2.7$\\
    & $42.9\pm7.7$ & $81.5\pm3.8$\\
    Willshaw & $86.9\pm7.0$ & $96.7\pm1.8$\\
    & $83.0\pm5.6$ & $95.9\pm1.6$\\
    & & \\
    \multicolumn{3}{l}{TKO} \\
    Experiment & $20.6\pm12.4$ & $64.9\pm13.7$ \\
    Gierer & $0.4\pm0.7$ & $20.0\pm5.4$\\
    Koulakov & $6.9\pm7.5$ & $38.9\pm12.8$\\
    Whitelaw & $0.1\pm0.3$ & $18.9\pm3.2$\\
    Willshaw & $25.9\pm21.2$ & $63.3\pm14.7$\\
    & & \\
    \multicolumn{3}{l}{\gMathV} \\
    Experiment & -  &  - \\
    Gierer & $27.9\pm8.4$ & $73.6\pm4.2$\\
    Koulakov & $77.2\pm8.8$ & $93.3\pm3.1$\\
    Whitelaw & $36.9\pm15.0$ & $76.5\pm7.3$\\
    Willshaw & $71.1\pm11.9$ & $91.0\pm4.5$\\
    \hline
  \end{tabular}
  \caption{Summary of Lattice measure for the largest locally ordered
    submap. The size is given both as the percentage of edges in the
    largest ordered submap, and as percentage of nodes in the largest
    ordered submap that have retained all their edges compared to the
    full map. Values are given
    as mean $\pm$ SD (N=10). Where experimental
    intrinsic imaging data is available \protect \cite{Cang2008}, the corresponding Lattice
    analysis values are reported \protect \cite{Willshaw2014}. For \gIslEkk and \gIslEkp the upper
    values are for the \gIslm map, and the lower values the \gIslp map. }
  \label{tab:lattice}
\end{table}

\clearpage

\begin{table}
  \begin{tabular}{ll}
  \hline
    & Retinal coverage \\
\hline
    Experiment (P8) & $11.1 \pm 9.1\,\%$ \\
    Experiment (P12) & $3.2 \pm 2.1\,\%$\\
    Experiment (P22) &  $2.6 \pm 1.1\,\%$ \vspace{2mm}\\
    Gierer & $6.9 \pm 2.0\,\%$ \\
    Koulakov & $4.0 \pm 1.0\,\%$ \\ 
    Whitelaw & $6.1 \pm 8.8\,\%$ \\
    Willshaw & $13.0 \pm 1.2\,\%$ \\
  \hline
  \end{tabular}
  \caption{Contour analysis of retinal labelling from 
    retrograde injections in the SC. Mean $\pm$ SD of retinal coverage for 95\,\% of the labeling
    are reported. Experimental data from \protect \citetext{Lyngholm2013}.}
  \label{tab:contour}
\end{table}

\clearpage

\begin{table}
{\renewcommand{\arraystretch}{1.4}
\begin{tabular}{lp{2.5cm}p{2.5cm}p{2.5cm}p{2.5cm}}
  \hline
  Genotype & Gierer & Koulakov & Whitelaw & Willshaw  \\
  \hline
  Wild type & \checkmark & \checkmark & \checkmark & \textbf{*} \checkmark
  \\
  \gIslEkk & \gIslp misfit & \gIslp misfit & \textbf{*}  \checkmark &
  \gIslp misfit \\
  \gIslEkp & No~collapse, \gIslp misfit & \textbf{*}
  \gIslp misfit & No~collapse, \gIslp
  misfit & No~collapse, \gIslp misfit \\
  TKO (no gradient) & No patches & Patches but no global order & No patches & Global~order but no polarity  \\
  TKO (weak gradient) & No patches & \checkmark & No patches &  Ordered map\\
  \gMathV & \textbf{*} \checkmark & \checkmark & Normal map & Normal map \\
  \hline
\end{tabular}
}
\caption{Summary of model evaluation. Asterisk (\textbf{*}) denotes which phenotype the
  model was optimised~for.
  \label{tab:summary}}
\end{table}

\end{document}